\documentclass[12pt,a4paper]{article}  
 \usepackage[skins,theorems]{tcolorbox}

\newtcolorbox{whitebox}{colback=white,colframe=black,boxrule=0.5mm,arc=4mm,auto outer arc}

\tcbset{highlight math style={enhanced,
  colframe=red,colback=white,arc=0pt,boxrule=1pt}}
  \usepackage[bookmarksopen, bookmarksnumbered, bookmarksopenlevel=2]{hyperref}
  \usepackage{tikz}
  \usepackage{tikz-3dplot}
  \usepackage{multirow}
 \usetikzlibrary{calc}
 \usetikzlibrary{decorations} %
 \usepackage[UKenglish]{babel}
 \usepackage[toc,page]{appendix}
 \usepackage{amsmath}
 \usepackage{amssymb}
 \usepackage{graphicx}
 \usepackage{hhline}
 \usepackage[bf]{caption}
\usepackage{cite}
\usepackage[vcentermath]{youngtab}
\usepackage{geometry}
\usepackage{slashed}
\usepackage{color}
\usepackage{stackrel}
\usepackage{tikz-cd} 
\usepackage{tkz-euclide}
\usepackage{cancel} 
\usepackage{subcaption}
\usepackage[normalem]{ulem}
\usepackage{mdframed}
\usepackage{adjustbox}
\usepackage{tcolorbox}
\usepackage{enumitem}
\usepackage{empheq}
\usepackage{arydshln}

\newenvironment{eqn*}{\begin{equation*}\begin{aligned}}{\end{aligned}\end{equation*}\noindent}

\newcommand{\bqa}{\begin{eqnarray}}
\newcommand{\eqa}{\end{eqnarray}}

\hypersetup{
    pdftitle={},
    pdfauthor={},
    pdfsubject={}
}
\numberwithin{equation}{section}
\numberwithin{table}{section}

\makeatletter

\definecolor{BF}{HTML}{f903d7}

\newtheorem{definition}{Definition}

\newcommand{\be}{\begin{equation}}
\newcommand{\ee}{\end{equation}}
\newcommand{\beq}{\begin{equation}}
\newcommand{\eeq}{\end{equation}}
\newcommand{\ba}{\begin{aligned}}
\newcommand{\ea}{\end{aligned}}

\newcommand{\bea}{\begin{eqnarray}}
\newcommand{\eea}{\end{eqnarray}}

\newcommand{\cO}{\mathcal{O}}

\newcommand{\cE}{\mathcal{E}}

\newcommand{\cN}{\mathcal{N}}

\newcommand{\cB}{\mathcal{B}}

\newcommand{\cU}{\mathcal{U}}

\newcommand{\cV}{\mathcal{V}}

\newcommand{\cM}{\mathcal M}

\newcommand\bi{\begin{itemize}}
\newcommand\ei{\end{itemize}}

\renewcommand{\a}{{\alpha}}

\def\Im{\mathop{\mathrm{Im}}\nolimits}

\def\unit{{1\kern-.65ex {\rm l}}}
\def\1{{1\kern-.65ex {\rm l}}}

\def\ii{{\rm i}}

\tcbset {
	base/.style={
		arc=0mm, 
		bottomtitle=0.5mm,
		boxrule=0mm,
		colbacktitle=black!10!white, 
		coltitle=black, 
		fonttitle=\bfseries, 
		left=2.5mm,
		leftrule=1mm,
		right=3.5mm,
		title={#1},
		toptitle=0.75mm,
		breakable
	}
}

\newtcolorbox{subbox}[1]{
	colframe=black!30!white,
	base={#1}
}

\newcount\hour \newcount\minute
\hour=\time \divide \hour by 60
\minute=\time
\def\now{%
\ifnum \hour<13
  \ifnum \hour=0 \advance \hour by 12 \number\hour:\else \number\hour:\fi%
     \ifnum \minute<10 0\fi%
     \number\minute%
\ A.M.%
\else \advance \hour by -12 \number\hour:%
  \ifnum \minute<10 0\fi%
  \number\minute%
  \ P.M.%
\fi%
}

\makeatother

\begin{document}
\begin{flushright}
 \small{ZMP-HH/26-4}
\end{flushright}

\vskip 40 pt
\begin{center}
{\large \bf
Quantum obstructions for $\cN=1$ infinite distance limits  -- \vspace{2mm}\\
 Part I: $g_s$ obstructions
} 

\vskip 11 mm

Lukas Kaufmann,  Jeroen Monnee, Timo Weigand, and Max Wiesner

\vskip 11 mm
\small \textit{II. Institut f\"ur Theoretische Physik, Universit\"at Hamburg, Notkestrasse 9,\\ 22607 Hamburg, Germany} \\[3 mm]
\small \textit{Zentrum f\"ur Mathematische Physik, Universit\"at Hamburg, Bundesstrasse 55, \\ 20146 Hamburg, Germany  }   \\[3 mm]

\end{center}

\vskip 7mm

\begin{abstract}

We analyse quantum obstructions to classical infinite distance limits in four-dimen\-sional string compactifications with ${\cal N}=1$ supersymmetry. Such quantum effects signal a severe departure from the perturbative effective action and can be of considerable importance for string model building.
Our focus is on the complex structure moduli space of  Type IIB orientifolds
with O7/O3-planes and its F-theory description.
 In this first part of our analysis, we investigate the behaviour of $g_s$ corrections in infinite distance complex structure limits. 
 Our main finding is that, depending on the location of the O7-plane, non-perturbative corrections in $g_s$ can become \textit{unsuppressed},  thus obstructing a perturbative Type IIB description in the corresponding asymptotic region of the field space. In particular, this applies to large complex structure limits.
 To show this, we study the F-theory description of the Type IIB orientifold, in which all pure $g_s$ corrections are encoded in the (classical) geometry of an elliptic Calabi--Yau fourfold. This $g_s$-corrected moduli space is found to differ significantly from the classical moduli space. In extreme cases the classical infinite distance degeneration can be completely removed at the $g_s$-corrected quantum level. The behaviour of $\alpha'$ corrections, as well as implications for string model building, are discussed in a companion paper. 
\end{abstract}

\vfill

\thispagestyle{empty}
\setcounter{page}{0}
 \newpage
\tableofcontents
\vspace{25pt} 
\setcounter{page}{1}

\section{Introduction}
Identifying a consistent theory of quantum gravity, capable of describing our universe at the fundamental level, is among the biggest challenges in theoretical physics.
A large class of consistent quantum gravity theories can be obtained from string theory and its compactifications. However, full computational control over these theories is typically only achieved in either highly supersymmetric setups or in the strict weak coupling limit. The reason is that it is primarily in such contexts that sufficiently many gravitational quantum effects such as string loop corrections are either suppressed (in the case of weak coupling) or forbidden by  non-renormalisation theorems (in presence of supersymmetry). Unfortunately, while the absence of certain quantum gravitational effects allows for computational tractability, this also goes hand-in-hand with the resulting theories being highly unrealistic such that they cannot describe our universe: Large amounts of supersymmetry forbid, for example, the existence of chiral fermions, while in strict weak-coupling regimes of quantum gravity it is difficult to obtain viable cosmological scenarios describing our observed universe~\cite{Dine:1985he}.

For these reasons, string theory based model building approaches typically rely on compactifications resulting in 4d $\cN=1$ theories of gravity where, at least in principle, the key requirements for a theory describing our universe can be met. The underlying logic is to maintain a similar degree of computational control as for theories with extended supersymmetry, but in a potentially more realistic setting. A common strategy to achieve this is to start with an effective 4d $\cN=2$ theory of gravity, and to subsequently break supersymmetry to 4d $\cN=1$. 

This is, in particular, the strategy behind Type II orientifold compactifications on Calabi--Yau threefolds. Concretely, Type IIB orientifolds are prominently featured in the KKLT~\cite{Kachru:2003aw} and LVS~\cite{Balasubramanian:2005zx,Conlon:2005ki} scenarios for moduli stabilisation. The general goal motivating this work and its companion paper \cite{Paper2} is to assess under what conditions the couplings of the resulting 4d $\cN=1$ effective theory can be reliably computed from the 4d $\cN=2$ parent theory.

In string compactifications, the couplings of the 4d effective actions  depend on the vacuum expectation values of the scalar fields in the theory. In a Type IIB compactification on an orientifold of a Calabi--Yau threefold $V$, the scalar fields descend as a subset of the moduli of the ${\cal N}=2$ parent theory.  The moduli space of this parent theory factors into two sectors, the hyper- and the vector multiplet moduli space, which are decoupled from each other. In particular, the vector multiplet moduli space can be described purely in terms of the complex structure deformations of $V$, since the string coupling $g_s$ resides in a hypermultiplet. For this reason, the couplings for the vector multiplets are exact and receive neither $g_s$ nor $\alpha'$ corrections. As a consequence, the classical effective action can be trusted even in regimes where either the string coupling becomes large or three-cycle volumes shrink to zero size. This includes limits in the complex structure moduli space of $V$ that lie at infinite distance, as they have been discussed in detail e.g. in~\cite{Grimm:2018ohb,Grimm:2018cpv,Joshi:2019nzi,Bastian:2020egp,Gendler:2020dfp,Palti:2021ubp,Grimm:2022sbl,Grimm:2022xmj,Hassfeld:2025uoy,Monnee:2025ynn,Monnee:2025msf,Hattab:2025aok,Hattab:2026lho}. In these limits, entire families of three-cycles can be shown to shrink to zero size~\cite{Hassfeld:2025uoy,Monnee:2025ynn,Monnee:2025msf}. In these regimes, the couplings of the vector multiplet sector of the 4d $\cN=2$ effective theory take a particularly simple form \cite{schmid, CKS,Candelas:1990rm,Candelas:1993dm,Candelas:1994hw,Bastian:2021eom,Bastian:2023shf}, and a subset of the gauge sector becomes weakly coupled. Based on this, one would expect the perturbative Type IIB description to be particularly well-suited to describe the effective physics in these regimes. 

However, it turns out that the perturbative Type IIB description is not the most adequate one in these extreme regimes of the vector multiplet moduli space. This is because there are infinite towers of states that asymptotically become light, as predicted by the Distance Conjecture~\cite{Ooguri:2006in}. These states are non-perturbative from the Type IIB string theory perspective, and furnish the degrees of freedom of a dual theory that emerges in the asymptotic limit. Determining the nature of the tower of states requires detailed knowledge about the geometry of the underlying Calabi--Yau threefold in the asymptotic regime which is not directly encoded in the low-energy effective action. The precise towers of states becoming light in the possible asymptotic regimes of the vector multiplet sector of Type IIB Calabi--Yau compactifications have been determined in~\cite{Hassfeld:2025uoy,Monnee:2025ynn,Monnee:2025msf}, with the results being in full accordance with the Emergent String Conjecture~\cite{Lee:2019oct}. To arrive at these results, the supersymmetric non-renormalisation theorem of the 4d $\cN=2$ effective theory was instrumental as it ensures that the physics of the vector multiplet sector as encoded in the geometry is not corrected by quantum effects. This is to be contrasted with classically infinite distance regimes in the hypermultiplet sector where perturbative and non-perturbative corrections considerably change the effective action~\cite{Marchesano:2019ifh,Baume:2019sry}.

Once half of the supersymmetries are broken through an orientifold action $\Omega$ on $V$, the non-renormalisation theorem  no longer holds, and the factorisation of the 4d $\cN=2$ moduli space into hyper- and vector multiplet sectors breaks down. As a result, quantum corrections that are ensured to be absent in $\cN=2$ theories correct the effective 4d $\cN=1$ action. Computing these corrections from first principles is a challenging task, see \cite{Becker:2002nn,vonGersdorff:2005bf,Berg:2005ja,Berg:2007wt,Berg:2011ij,Grimm:2013gma,Grimm:2013bha,Berg:2014ama,Minasian:2015bxa,Weissenbacher:2019mef,Klaewer:2020lfg,Cicoli:2021rub,Wiesner:2022qys,Cvetic:2024wsj,Casas:2025qxz} for an incomplete collection of results for perturbative and non-perturbative corrections to the classical 4d $\cN=1$ effective action. As a consequence, the string coupling does not have to decouple from the scalar fields that descend from the vector multiplets in the $\cN=2$ parent theory. In this work, we focus on Type IIB orientifold actions $\Omega$ that lead to O7/O3-planes. In this case, the complex structure moduli of the original Calabi--Yau threefold $V$ that survive the orientifold projection correspond to the elements of $H^3_-(V/\Omega)$, i.e., harmonic 3-forms that are anti-invariant under the orientifold $\mathbb{Z}_2$-involution. At tree-level, the effective action of the resulting 4d $\cN=1$ theory is obtained as a certain projection of the 4d $\cN=2$ effective action, see~\cite{Grimm:2004uq,Grimm:2005fa}. Focusing on the scalar modes associated with complex structure deformations of the Calabi--Yau threefold, the tree-level couplings are still independent of the string coupling. This reflects the factorisation of the moduli space in the 4d $\cN=2$ parent theory. In the 4d $\cN=1$ theory, perturbative and non-perturbative effects can correct the effective action in the complex structure sector. If present, such corrections are genuine $\cN=1$ effects, for which a direct computation within perturbative Type IIB string theory is rather challenging. To avoid these computational difficulties, one could focus on classical weak-coupling limits such as those arising in the asymptotic regimes of the complex structure moduli space, where the couplings of the effective action take a particularly simple form. This approach is commonly pursued in concrete model building approaches based on Calabi--Yau threefolds with the asymptotic regime corresponding, e.g., to the Large Complex Structure (LCS) regime. 

From the low-energy perspective one might expect that asymptotic regimes in the complex structure moduli space, for example the LCS regime, are under perfect computational control even at finite $g_s$ -- after all, these correspond to weak coupling limits of a dual theory. However, as we reviewed above, already in the parent 4d $\cN=2$ theory these limits are highly non-perturbative from the Type IIB perspective. This is due to the appearance of a tower of states that is of non-perturbative nature. In the parent theory, the extended supersymmetry ensures that the low-energy effective action does not receive corrections beyond tree-level. This is no longer the case after orientifolding. Due to the non-perturbative nature of the asymptotic regimes, the (non-)perturbative corrections to the perturbative Type IIB effective action can even become particularly strong in these regimes. Thus, instead of having excellent computational control over the effective action in the asymptotic regimes in the complex structure moduli space, the opposite may be true! 

The analysis of this work and the companion paper \cite{Paper2} shows that this is indeed the case, i.e., that there are asymptotic regimes in the complex structure moduli of Calabi--Yau orientifolds that cannot be described using the perturbative Type IIB effective action. As we will see, this in particular applies to the LCS regime. Whether or not the effective action in the complex structure sector receives strong $g_s$ corrections depends on the geometric properties of the orientifold action. This information goes beyond what is straightforwardly visible from the effective action, i.e., beyond the action of the orientifold on $H^3(V)$. To determine the effect of $g_s$ corrections in the 4d $\cN=1$ effective action, the crucial information concerns the fixed locus of the orientifold action relative to the degeneration of $V$ in the asymptotic regime. As we will show, there are two different types of orientifolds for a given limit in the complex structure moduli space which we dub O-type A and O-type B. See Figure \ref{fig:O-types} for an illustration. We find that in O-type A limits, $g_s$ corrections become unsuppressed in the infinite distance limit, whereas in O-type B models, they remain under perturbative control. 

As is well-known, the effect of $g_s$ corrections to the Type IIB effective action are encoded in the geometry of the F-theory lift~\cite{Vafa:1996xn,Sen:1996vd,Billo:2010mg}. In F-theory, the Calabi--Yau orientifold is replaced by an elliptically fibered Calabi--Yau fourfold. Its complex structure moduli space encodes the $g_s$-corrected Type IIB moduli space including the axio-dilaton, the complex structure moduli of $V$ surviving the orientifold projection as well as the 7-brane moduli. The complex structure moduli spaces of concrete examples of Calabi--Yau fourfolds have been discussed e.g. in~\cite{Greene:1993vm,Mayr:1996sh,Klemm:1996ts,Jockers:2009ti,Intriligator:2012ue,Gerhardus:2016iot,Cota:2017aal,Grimm:2019ixq,vandeHeisteeg:2024lsa,Grimm:2024fip}. In particular, the F-theory uplift of the D-brane superpotential and corresponding moduli~\cite{Lerche:1998nx,Lerche:2001cw,Lerche:2002ck,Jockers:2008pe,Jockers:2009mn} has been analysed in~\cite{Alim:2009rf,Alim:2009bx,Grimm:2009ef,Alim:2010za}. As we will see, in the F-theory lift the difference between the O-type A and B limits is manifest in the degeneration properties of the base threefold. For O-type~B, the degeneration of the base mimics the degeneration of the original Calabi--Yau threefold, whereas for O-type A the singularity can change drastically. By directly studying the fourfold complex structure moduli space we show that the $g_s$-corrected Type IIB moduli space differs considerably from the classical moduli space, implying that $g_s$ corrections become unsuppressed in the asymptotic limits in the complex structure moduli space. Most strikingly, we find that this is even true in the strict $g_s\to 0$ limit. The reason is that even in the strict Sen-limit for the F-theory fourfold~\cite{Sen:1996vd}, the infinite distance limit in the complex structure moduli space forces two O7-plane strata to coincide, which is a highly non-perturbative configuration from the Type IIB perspective. See \cite{Lee:2021qkx,Lee:2021usk} for a discussion of this effect in F-theory compactifications on elliptic K3 surfaces. As a consequence, the perturbative Type IIB moduli space is only a good approximation to the actual quantum moduli space in the strict weak coupling limit $g_s\to 0$ if no additional limit in the complex structure sector is taken. As we will see, in extreme cases the naive, classical infinite distance limit is even completely removed at the $g_s$-corrected quantum level.

This paper is structured as follows: In Section~\ref{sec:overview}, we give an overview of our main results and discuss a simple example where the effects of orientifolding on the effective action become clear. In Section~\ref{sec:IIBorientifolds}, we review some required background on Type IIB orientifolds and discuss the detailed geometries arising at infinite distance in the complex structure moduli space of Calabi--Yau threefolds before and after orientifolding. We further introduce the two types of orientifolds, O-type A and O-type B, and discuss their F-theory uplifts. We discuss the structure of the $g_s$-corrected Type IIB moduli space in Section~\ref{sec:p-vs-np}. In particular, we show how it differs from the classical Type IIB moduli space. We conclude in Section~\ref{sec:conclusions} and relegate details on the main examples considered in this work to the appendices.

\section{General overview and a simple example}\label{sec:overview}

 This work focuses on infinite distance loci in the complex structure moduli space of Calabi--Yau orientifold compactifications of Type IIB string theory. As explained in the introduction, we are interested in the question whether the classical effective action derived from perturbative Type IIB string theory remains under perturbative control in these asymptotic regions. We will find that generically, the classical effective action is not under perturbative control and $g_s$ corrections can significantly change the physics in the asymptotic regions of moduli space. The analogous effect of quantum corrections associated with K\"ahler moduli is the topic of the companion paper \cite{Paper2}.

An overview of the results of the analysis in this paper is provided in Section~\ref{ssec:gsoverview}.
The main lines of arguments are illustrated in a simple warm-up example in Section~\ref{ssec:simple-ex} based on a Type IIB orientifold of K3$\times T^2$.

\subsection{F-Theory and \texorpdfstring{$g_s$}{gs} corrections}\label{ssec:gsoverview}

Type IIB orientifolds of Calabi--Yau threefold compactifications form a well-studied class of 4d $\cN=1$ theories.  These constructions play a prominent role in particular in string theory model building for cosmology, see e.g. the reviews \cite{Cicoli:2023opf,McAllister:2025qwq} and references therein. For definiteness, we consider orientifold actions that are compatible with O7- and O3-planes. The classical 4d $\cN=1$ effective action obtained for such orientifolds has been analysed in detail in~\cite{Grimm:2004uq,Grimm:2005fa} (see also, e.g., \cite{Blumenhagen:2006ci,Ibanez:2012zz} for further background). As mentioned in the introduction, one of the main appeals of using Type IIB orientifolds for model building is that the classical effective action resembles a theory with extended $\cN=2$ supersymmetry since, at this level, the orientifold action simply projects onto a subsector of the 4d $\cN=2$ effective action obtained from Type IIB string theory compactified on a Calabi--Yau threefold $V$. For this 4d $\cN=2$ parent theory, the additional supersymmetry ensures a factorisation of the moduli space into 
\begin{equation}\label{eq:MN2}
    \cM_{\cN=2}(V) = \cM_{\rm HM}(V) \times \cM_{\rm VM}(V)\,. 
\end{equation}
Here, the first factor corresponds to the scalar fields in the hypermultiplet sector which, in the case of Calabi--Yau threefold compactifications of Type IIB string theory, is spanned by the dilaton, the complexified K\"ahler moduli of $V$ and the corresponding RR-axions. The second factor is spanned by the scalars in 4d $\cN=2$ vector multiplets which, in the Type IIB context, correspond to the complex structure deformations of $V$. At string tree-level, the couplings in the vector multiplet sector of Type IIB compactified on $V$ can be entirely described by the periods of the holomorphic three-form $\Omega_3$ on $V$. Due to the factorisation of $\cM_{\cN=2}(V)$ and the fact that the string coupling is part of the hypermultiplet sector, the vector multiplet moduli space is tree-level-exact in Type IIB compactifications on Calabi--Yau threefolds.  

Once half of the supersymmetries are broken by an orientifold action $\Omega$, the closed string part of the moduli space $\cM_{\cN=1}(V/\Omega)$ is classically a subset of $\cM_{\cN=2}(V)$. However, the non-renormalisation theorem underlying the factorisation~\eqref{eq:MN2} ceases to hold such that, at the quantum level, $\cM_{\cN=1}(V/\Omega)$ does not factorise into multiple components. Nonetheless, at small coupling and large volume of $V$, the 4d $\cN=1$ moduli space is expected to factorise approximately as 
\begin{equation}\label{eq:MN1VOmega}
    \cM_{\cN=1}(V/\Omega)\big\vert_{g_s=0\,,\;\text{vol}(V/\Omega) M_s^6 =\infty} \approx \cM_{\rm c.s.}(V/\Omega) \times \cM_\tau \times \cM_{\rm open}\times \cM_{\rm K}(V/\Omega)\,.
\end{equation}
Here, $\cM_{\rm c.s.}(V/\Omega)$ denotes the complex structure deformations of $V$ that are not projected out by $\Omega$, $\cM_\tau$ is spanned by the Type IIB axio-dilaton, and $\cM_{\rm K}(V/\Omega)$ is spanned by the K\"ahler deformations and RR-axions compatible with $\Omega$. Finally, $\cM_{\rm open}$ is the open string moduli space describing the position moduli of 7-branes.\footnote{If there are spacetime-filling D3-branes, the moduli space has an additional factor  corresponding to their position moduli. We do not discuss these moduli in this work.} 

In this paper, we first discuss finite string coupling effects, postponing additional finite volume effects to~\cite{Paper2}. We therefore formally work at infinite string-frame volume of $V/\Omega$ and study to what extent $g_s$ corrections spoil the factorisation between the first three factors in~\eqref{eq:MN1VOmega}. In particular, we ask whether the classical effective action remains under perturbative control as we approach classical infinite distance limits in $\cM_{\rm c.s.}(V/\Omega)$. In other words, we consider the $g_s$-corrected moduli space at infinite volume, which takes the form 
\begin{equation}\label{eq:MN1Vgs}
    \cM_{\cN=1}(V/\Omega)\big\vert_{\text{vol}(V/\Omega) M_s^6 =\infty} = \cM_{Q}(V/\Omega) \times \cM_{\rm K}(V/\Omega)\,, 
\end{equation}
and analyse if infinite distance regimes of $\cM_{\rm c.s.}(V/\Omega)$ lift to infinite distance regimes of $\cM_Q(V/\Omega)$. As it turns out, this is not always the case. Instead, we will find that certain infinite distance limits in $\cM_{\rm c.s.}(V/\Omega)$ are obstructed in $\cM_Q(V/\Omega)$ in the following sense:  

\begin{definition}\label{def:2fusionheunconnected}
    Consider Type IIB string theory compactified on an orientifold of a Calabi--Yau threefold $V$ with orientifold action $\Omega$ leading to O7/O3-planes and consider an infinite distance limit in the complex structure moduli space $\cM_{\rm c.s.}(V/\Omega)$. Such a limit is called $g_s{\rm\text{-}obstructed}$ if in the $g_s$-corrected moduli space $\cM_Q(V/\Omega)$ the following two conditions cannot be satisfied simultaneously in the limit:
    \begin{enumerate}
        \item The Type IIB string coupling remains finite, $g_s>0$, along the limit. 
        \item The 4d $\cN=1$ effective action derived from Type IIB string theory on $V/\Omega$ remains under perturbative control in $g_s$. 
    \end{enumerate}
\end{definition}
To test whether an infinite distance limit in $\cM_{\rm c.s.}(V/\Omega)$ is $g_s$-obstructed, we must know the $g_s$-exact quantum moduli space $\cM_Q(V/\Omega)$. To this end, we consider the F-theory uplift of the Type IIB orientifold $V/\Omega$, which completely geometrises all pure $g_s$ corrections.\footnote{By \emph{pure} $g_s$ corrections we mean those that are not mixed $g_s$, $\alpha'$ corrections to the Type IIB effective action.} In F-theory, the Type IIB orientifold is replaced by an elliptically fibered Calabi--Yau fourfold 
\begin{equation}
\cE \hookrightarrow W \to \cB_3 \,,
\end{equation}
where $V$ is a double cover of $\cB_3$. The $g_s$-exact moduli space can then be identified with the complex structure moduli space of $W$,
\begin{equation} 
    \cM_Q(V/\Omega) = \cM_{\rm c.s.}(W)\,. 
\end{equation} In particular, the Type IIB string coupling is identified with a complex structure modulus of $W$ such that the weak coupling limit $g_s\to 0$ corresponds to the Sen-limit \cite{Sen:1996vd}, in which the fiber $\cE$ of $W$ degenerates over generic points on $\cB_3$~\cite{Clingher:2012rg}. 
 In this limit, the degenerate elliptic fibration develops a bi-section which can be identified with the original Calabi--Yau threefold $V$. Away from the Sen-limit, the O7-planes appearing in $V/\Omega$ are resolved in the F-theory lift, implying that the geometry of $W$ differs considerably from that of a trivial fibration $V/\Omega \times T^2$. As we will see in detail in Section~\ref{sec:IIBorientifolds}, this distinction can make a significant difference in the vicinity of infinite distance limits in $\cM_{\rm c.s.}(V/\Omega)$. From the perspective of $V$, such limits correspond to degenerations of the Calabi--Yau threefold. If the O7-planes lie in a specific way inside this degenerate geometry, which will be made precise in Section~\ref{ssec:oriendegenerations}, the geometry of the uplift $W$ considerably differs from the classical geometry of $V/\Omega$. As a consequence, the effective action derived from Type IIB on $V/\Omega$ does not capture the physics in this regime correctly. 
 
 In fact, we will see that 
 the effective action differs even for $g_s \to 0$ if at the same time we approach the infinite distance regime in $\cM_{\rm c.s.}(V/\Omega)$. This is a very drastic effect which can be traced back to a non-perturbative factorisation of the O7-planes in the limit. 
 The limit is then $g_s$-obstructed in the sense of Definition~\ref{def:2fusionheunconnected}.
 We will present the details of how F-theory deals with the classical infinite distance limits in $\cM_{\rm c.s.}(V/\Omega)$ in Section~\ref{sec:p-vs-np}.

\subsection{A simple example: \texorpdfstring{$g_s$}{gs} obstructions in O-Type A orientifolds}\label{ssec:simple-ex}
To illustrate the notion of $g_s$-obstructions in a concrete yet accessible example, consider an orientifold compactification of Type IIB string theory on the Calabi--Yau threefold $V={\rm K3}\times T^2$. The extended $\cN=2$ supersymmetry after orientifolding makes an explicit analysis of instanton corrections tractable without having to invoke F-theory. Of particular interest to us is the vector multiplet sector of the resulting 4d $\cN=2$ theory. We focus on the subsector of this moduli space that is spanned by the three moduli $S, T, U$ which can be viewed, respectively, as the moduli associated with the string coupling, a K\"ahler modulus, and a complex structure modulus. Classically, this slice of the moduli space is a product manifold 
\begin{equation}\label{MK3timest2}
    \cM_{\rm VM,cl}(V/\Omega)\supset\cM_{S} \times \cM_{T} \times \cM_U\,,
\end{equation}
in which $S$, $T$, and $U$ can be varied independently. This classical moduli space should be viewed as the K3$\times T^2$ analogue of~\ref{eq:MN1VOmega}. Classically, each of the factors in \eqref{MK3timest2} is a copy of ${\rm SL}(2,\mathbb{Z})\backslash {\rm SL}(2,\mathbb{R})/{\rm U}(1)$ such that there are infinite distance limits in each of the factors. The factorisation, however, does not hold at the quantum level since quantum corrections in $g_s$ and $\alpha'$ non-trivially affect the geometry of the quantum vector multiplet moduli space $\cM_{{\rm VM},Q}(V/\Omega)$. For this reason, the infinite distance limits in each of the factors appearing in~\eqref{MK3timest2} may not be realized in $\cM_{{\rm VM},Q}(V/\Omega)$. Here, we are particularly interested in the fate of the infinite distance limit in $\cM_U$ corresponding to $U\to \ii \infty$: $U$ parametrises the complex structure of the $T^2$ factor such that $\cM_U$ plays the role of $\cM_{\rm c.s.}(V/\Omega)$ in the general discussion above. 

 For a certain choice of the orientifold action, we argue now that the $U\to \ii \infty$ limit is $g_s$-obstructed in the sense of Definition~\ref{def:2fusionheunconnected} (while for the other choice, an analogous obstruction due to $\alpha'$-effects will be found in \cite{Paper2}). Let us stress that this example illustrates that $g_s$-obstructions are features that are not exclusive to 4d $\cN=1$ theories, but arise whenever the complex structure moduli space does not decouple from the string coupling as would be the case if the moduli space is a direct product also at the quantum level.\\

For $V={\rm K3}\times T^2$ there are two projections leading to orientifolds with O7-planes:\footnote{For general Calabi--Yau threefolds, the notion of {\it O-type} will be defined in Section~\ref{ssec:oriendegenerations} and the examples presented here are indeed examples of O-type A and B orientifolds with respect to the large complex structure limit of the $T^2$.}
\begin{itemize}
    \item\textbf{O-type A:} $\Omega_{\rm A}=\Omega_p(-1)^{F_L} R$, where $\Omega_p$ denotes worldsheet parity, $(-1)^{F_L}$ is the left-moving fermion number and $R$ acts on the $T^2$-coordinate $z$ by reflection, $R:z\mapsto-z$. This reflection has $4$ fixed points on $T^2$, thereby signalling the presence of $4$ O7-planes wrapping the K3 in the orientifold.
    \item\textbf{O-type B:} $\Omega_{\rm B}=\Omega_p(-1)^{F_L}\rho$, where $\rho$ is an anti-symplectic involution on the ${\rm K3}$, i.e. $\rho^\ast(\omega_{\rm K3})=-\omega_{\rm K3}$. The fixed point locus of $\rho$ consists of a union of curves $C_I$ inside the ${\rm K3}$, corresponding to O7-planes wrapping $C_I\times T^2$.
\end{itemize}
In the following, we discuss the O-type A orientifold and show that the large complex structure limit of the $T^2$ is obstructed in the sense of Definition~\ref{def:2fusionheunconnected} by studying the action of D$(-1)$-instantons in this limit. In~\cite{Paper2} we will find an analogous obstruction for the orientifold of O-type B due to $\alpha'$ effects.

For the O-type A orientifold, the relevant moduli in the vector multiplet sector are
\begin{equation}
    \tau=C_0+\frac{\ii}{g_s}\,,\quad T_{\rm K3}=\int_{\rm K3}\left(C_4+\frac{\ii}{2}J^2\right)\,,\quad U\,,
\end{equation}
where $\tau$ is the Type IIB axio-dilaton, $J$ is the K\"ahler form on K3 and $U$ is the complex structure of $T^2$. The type II degeneration of $V$ described above then is $U\to\ii\infty$.

To establish that this limit is $g_s$-obstructed in the sense of Definition~\ref{def:2fusionheunconnected}, we consider corrections to the $\cN=2$ K\"ahler potential due to D$(-1)$-instantons. By performing two T-dualities along the $T^2$, the O-type A orientifold with O7-planes is mapped to Type I string theory with O9-planes wrapping the full internal space ${\rm K3}\times T^2$. A special instance of such a model is the Bianchi--Sagnotti--Gimon--Polchinski orbifold~\cite{Bianchi:1990tb,Gimon:1996rq} for which the K3 is realized at its orbifold point $T^4/\mathbb{Z}_2$.\footnote{The K3-moduli are part of the hypermultiplet sector such that for our analysis of the vector multiplet sector, we can focus w.l.o.g. on the orbifold point in the K3 moduli space. At this point additional O5-planes wrapping the $T^2$-factor of $V$ arise at the fixed points of the $\mathbb{Z}_2$-orbifold action.}
In Type I language, the relevant physical moduli at tree-level are given by
\begin{equation}
    S=\int_{{\rm K3}\times T^2}C_6+\ii\left(e^{-2\phi}{\rm vol}({\rm K3}\times T^2)M_s^6\right)\,, \quad T=\int_{T^2}C_2+\ii\left(e^{-\phi}{\rm vol}(T^2)M_s^2\right)\,,\quad U_I\,,
\end{equation}
where $\phi$ is the 10d Type IIB dilaton, $M_s$ the 10d string scale and $U_I$ is the complex structure of the torus. The three complex moduli $(S,T,U_I)$ form the scalar components of three vector multiplets that are T-dual to the O-type A vector multiplets mentioned above. More precisely, T-duality identifies
\begin{equation}
    \tau\,\leftrightarrow\,T,\quad T_{\rm K3}\,\leftrightarrow\,S,\quad U\,\leftrightarrow\,-\frac{1}{U_I}\,.
    \label{eq:T-dual-IIB_I}
\end{equation}
Non-perturbative corrections to the Type I K\"ahler potential have been computed in~\cite{Camara:2008zk} using duality to the heterotic string. The classical K\"ahler potential on the Type I vector multiplet moduli space is given by
\begin{equation}
    K_{\rm cl}=\log\left[(S-\bar{S})(T-\bar{T})(U_I-\bar{U}_I)\right]\,,
\end{equation}
which receives corrections both at the perturbative and the non-perturbative level. The analysis of~\cite{Camara:2008zk} takes into account the perturbative as well as the D1-instanton contribution yielding the corrected K\"ahler potential
\begin{equation}\label{eq:K-corr-TypeI}
    K=-\log\left[(S-\bar{S})(T-\bar{T})(U_I-\bar{U}_I)\right]+\frac{\ii}{2}\frac{V_{\rm 1-loop}+V_{\rm D1}}{S-\bar{S}}\,.
\end{equation}
In Appendix~\ref{app:TypeI-D1} we collect the explicit expressions for $V_{\rm 1-loop}$ and $V_{\rm D1}$ as well as details on the large $\Im(U_I)$-expansion of the latter. The result of this analysis is the expression
\begin{equation}\label{eq:VD1-largeU}
    V_{\rm D1}=-\frac{1}{\pi}\sum_{p=1}^\infty\frac{e^{2\pi\ii p T}}{p^2}\left(\frac{1}{{\rm Im}(T)}\left(\frac{1}{2}-\frac{1}{4\pi p {\rm Im}(U_I)}\right)-\frac{1}{{\rm Im}(U_I)}\right)e^{-2\pi\ii U_I}+\cO(1)+{\rm c.c.}\,.
\end{equation}
Notice that this series only converges for ${\rm Im}(U_I)<{\rm Im}(T)$, which is due to the sign of $U_I$ in the leading exponent. 

T-dualising back to the O-type A orientifold of Type IIB string theory, the identifications~\eqref{eq:T-dual-IIB_I} show that the D1-instantons in Type I map to D$(-1)$-instantons on the Type IIB side. The O-type A K\"ahler potential is hence corrected as in~\eqref{eq:K-corr-TypeI} with $V_{\rm D1}$ now giving the contribution of the D$(-1)$-instantons. As $V_{\rm D1}$ is a modular function, see~\eqref{eq:D1-modular}, and T-duality acts on the complex structure of the torus as in~\eqref{eq:T-dual-IIB_I}, the instanton correction in the limit $U\to\ii\infty$ behaves in the same way as in the limit $U_I\to\ii\infty$ as computed above. Therefore, the D$(-1)$-instantons in O-type A become unsuppressed, showing that the naive complex structure limit $U\to\ii\infty$ cannot be taken without losing perturbative control in $g_s$. 

In order to maintain perturbative control (at least in $g_s$), it is necessary to superimpose the limit $U_I\to\ii\infty$ with
\begin{equation}\label{eq:K3T2-coscaling}
    \frac{1}{g_s}\sim{\rm Im}(U)\to\infty\,,
\end{equation}
so that the corrected D$(-1)$-instanton action 
\begin{equation}
    S_{\rm D({\text-}1)}^{\rm corr} = \tau - U
    \label{eq:SD(-1)-corr}
\end{equation}
stays finite. Comparing this situation to Definition~\ref{def:2fusionheunconnected}, we conclude that the O-type A limit $U\to\ii\infty$ in this Type IIB orientifold is $g_s$-obstructed.

Note that from the point of view of the effective action, the threshold correction in \eqref{eq:SD(-1)-corr}
 can be absorbed by a field redefinition. The pathology of the original limit $U \to \ii \infty$ is reflected in the fact that the  redefined modulus (controlling the BPS instanton action)
 would tend to zero.

The co-scaling~\eqref{eq:K3T2-coscaling} is in fact only necessary and not sufficient to completely cure the limit. This is for two reasons. First, as can be seen by further duality to a Type IIA CY$_3$-compactification, there will be (at least) further corrections due to D$3$-brane instantons. Details on this are provided in Appendix~\ref{app:TypeI-D1}. Second, as will be discussed in detail in Section~\ref{sec:p-vs-np}, the co-scaling~\eqref{eq:K3T2-coscaling} (more precisely, $g_s^{-1}\gg\Im U\to\infty$) can be seen to lead to a collision of O7-planes, making it highly non-perturbative from the IIB perspective such that the perturbative Type IIB effective action cannot describe this limit.

\section{Complex structure limits in Type IIB orientifolds}\label{sec:IIBorientifolds}

In this section we analyse the geometry of Calabi--Yau orientifolds in asymptotic regimes of their complex structure moduli space and discuss their F-theory uplift. 
After reviewing the effective action of Type IIB orientifolds and the geometry of complex structure deformations in the $\cN=2$ parent theories in Sections~\ref{ssec:N=1action} and~\ref{ssec:CY3deformations}, respectively, we discuss the effect of the orientifold on these classical infinite distance limits in Section~\ref{ssec:oriendegenerations}. 
 A first step to address the fate of these limits in the physical, fully quantum-corrected moduli space is by considering their lift to F-theory, which automatically incorporates the perturbative and non-perturbative corrections in $g_s$. 
In Section~\ref{ssec:Ftheorylift}, we introduce the main example for our analysis. In particular, in this example we can identify a limit that is at infinite distance in the Type IIB moduli space, but turns into a finite distance limit when lifted to F-theory. A more general discussion and the physical interpretation of the geometries analysed here is provided in Section~\ref{sec:p-vs-np}.

\subsection{Review: 4d \texorpdfstring{$\cN=1$}{N=1} effective action}\label{ssec:N=1action}

Some essential properties of the four-dimensional ${\cal N}=2$ effective action of Type IIB string theory compactified on a Calabi--Yau threefold $V$ have already been recalled in Section \ref{ssec:gsoverview}, and we review additional details here.
 Apart from the 4d $\cN=2$ gravity multiplet, the theory contains $h^{2,1}(V)$ vector multiplets and $h^{1,1}(V)+1$ hypermultiplets. While the universal hypermultiplet contains as its bosonic components the Type IIB axio-dilaton $\tau=C_0+\frac{\ii}{g_s}$ together with the periods of $C_6$ and $B_6$ over $V$, the scalar components of the remaining $h^{1,1}(V)$ hypermultiplets are made up of the K\"ahler moduli of $V$ together with the periods of $B_2$ and $C_2$ along curves in $V$ and the periods of $C_4$ along divisors of $V$. Altogether, the scalars in the hypermultiplets form a quaternionic K\"ahler manifold of real dimensions $4h^{1,1}(V)+4$. In contrast, the vector multiplets contain vector fields that arise by expanding the Type IIB 4-form potential $C_4$ in a basis of three-forms, 
\begin{equation}
    C_4 = A^i \wedge \alpha_i + B_i \wedge \beta^i \,,\quad \alpha_i,\beta^i \in H^3(V)\,, \quad  \int_V \alpha_i \wedge \beta^j=\delta_i^j\,. 
\end{equation}
Accordingly, in a given basis of $H^3(V)$ there are $h^{2,1}(V)+1$ electric vectors out of which $h^{2,1}(V)$ are the vector components of the vector multiplets. The additional vector field is identified with the graviphoton in the gravity multiplet. The scalar fields in the vector multiplets correspond to the complex structure moduli of $V$. The complex structure moduli space $\cM_{\rm c.s.}(V)$ of $V$ has complex dimension $h^{2,1}(V)$ and is parametrised by complex coordinates $u^i$. It is a special K\"ahler manifold for which the K\"ahler potential is given by 
\begin{equation}
    K_{\rm c.s.} = - \log \int_V \Omega_3 \wedge \bar{\Omega}_3\,.
\end{equation}
The $u^i$-dependence of the K\"ahler potential enters through the holomorphic $(3,0)$-form $\Omega_3$ on $V$. In a local patch around a point $a^i\in \cM_{\rm c.s.}(V)$, we can define flat coordinates on the moduli space as
\begin{equation}
    z^i = \frac{1}{2\pi \ii}\log (u^i-a^i)\,. 
\end{equation}
The kinetic terms for the scalar fields $z^i$ are then given by the K\"ahler metric on $\cM_{\rm c.s.}(V)$,
\begin{equation}\label{eq:metricVM}
    g_{i\bar{\jmath}} = \partial_{z^i} \partial_{\bar{z}^j} K_{\rm c.s.} \,. 
\end{equation}
Due to the $\cN=2$ supersymmetry, the vector- and hypermultiplet sectors completely decouple such that the field space metric for the vector multiplet moduli is independent of the scalars in the hypermultiplets and vice versa. An important consequence of this decoupling is that the metric~\eqref{eq:metricVM} on the 4d $\cN=2$ vector multiplet moduli space is classically exact and does not receive any quantum corrections.

Starting from the 4d $\cN=2$ theory obtained from the Calabi--Yau threefold compactification of Type IIB string theory, we can obtain a 4d $\cN=1$ theory through an orientifold projection. In this work, we focus on Type IIB orientifolds with O7- and O3-planes. To this end, we consider an orientifold action $(-1)^{F_L}\Omega_p \sigma$ where $F_L$ is the left-moving spacetime fermion number, $\Omega_p$ is worldsheet parity and $\sigma$ is a geometric holomorphic involution on the Calabi--Yau threefold $V$. As a consequence of the orientifold projection, some of the massless degrees of freedom are projected out and hence become massive. All scalars surviving the orientifold projection sit in 4d $\cN=1$ chiral multiplets. In addition, due to the presence of 7- and 3-branes there are open string moduli. Classically, these can be treated separately from the closed string moduli on which we focus here. With respect to the orientifold action, the space $H^3(V)$ splits as $H^3_+(V)\oplus H^3_-(V)$. Accordingly, the basis $(\alpha_i,\beta^i)$ splits into a basis $(\alpha_\kappa,\beta^\kappa)$ of $H^3_+(V)$ and $(\alpha_{\hat k}, \beta^{\hat{k}})$ of $H^3_-(V)$. Since the holomorphic 3-form is anti-invariant under $\sigma$, it can be expanded as
\begin{equation}
    \Omega_3 = Z^{\hat k} \alpha_{\hat k} - F_{\hat k} \beta^{\hat k}\,, 
\end{equation}
such that after orientifolding, the K\"ahler potential for the complex structure sector is given by
\begin{equation}
    K_{\rm c.s.} = -\log \int_V \Omega_3 \wedge \bar{\Omega}_3 = -\log\left(Z^{\hat k}\bar{F}_{\hat k}- F_{\hat k} \bar{Z}^{\hat{k}}\right)\,. 
\end{equation}
In addition, each hypermultiplet of the parent 4d $\cN=2$ theory gives rise to one chiral multiplet. Under $\sigma$, the second cohomology of $V$ splits as $H^2(V) = H^2_+(V)\oplus H^2_-(V)$. Since $B_2$ and $C_2$ are odd under $(-1)^{F_L}\Omega_p$, these give rise to the scalar modes in $h^2_-(V)$ chiral fields. In the following, we ignore these purely axionic chiral multiplets as they do not play any role for the study of infinite distance limits. Instead, we focus on the $h^2_+(V)$ chiral multiplets with scalar components given by
\begin{equation}
    T_a = \frac12 \int_{D_a} (C_4 + \frac{\ii}{g_s} J^2) \,,\qquad D_a \in H_4^+(V)\,. 
\end{equation}
In addition, we obtain a chiral multiplet from the universal hypermultiplet with scalar component given by $\tau$. At leading order in $g_s$ and $\alpha'$, the K\"ahler potential for the closed string moduli then takes the form 
\begin{equation}
    K = K_{\rm c.s.}(z) + K_{\rm K}(\tau, T) = -\log\left[\int_V \Omega_3 \wedge \bar\Omega_3 \right] - \log(\tau-\bar\tau) - 2 \log(\cV_E(V)) \,,
\end{equation}
where 
\begin{equation}
    \cV_E(V) = \frac{1}{g_s^{3/2}} \int_V J^3\,
\end{equation}
is the 10d Einstein frame volume of $V$. Notice in particular that, at this classical level, the 4d $\cN=1$ field space still factorises into four sectors corresponding to the complex structure moduli, the axio-dilaton, the K\"ahler moduli and the open string moduli space, which we did not discuss in detail here.

\subsection{Review: Geometry of Calabi--Yau threefold degenerations}\label{ssec:CY3deformations}
\begin{figure}
    \centering
    \includegraphics[width=0.6\linewidth]{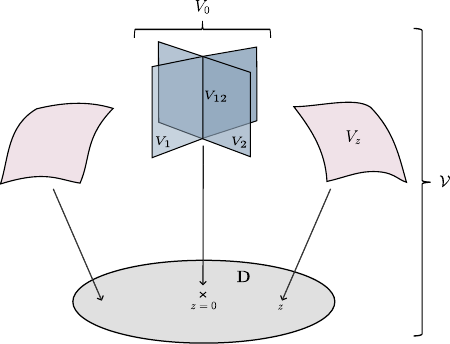}
    \caption{A semi-stable degeneration as in \eqref{eq:3foldV} in which the central fiber $V_0$ splits into the union of two components $V_1$ and $V_2$ that intersect over $V_{12}$.}
    \label{fig:ss_degeneration}
\end{figure}

In this section, we review the geometric interpretation of infinite distance limits in the complex structure moduli space of (unorientifolded) Calabi--Yau threefolds, as discussed in more detail in \cite{Monnee:2025ynn}. In particular, we introduce the classification of such limits into types II, III, and IV for semi-stable degenerations, based on the intersection properties of the various components into which the Calabi--Yau degenerates; see~\cite{Grimm:2018ohb,Grimm:2018cpv} for the Hodge-theoretic perspective on this classification. To this end, we consider a family \begin{equation}\begin{aligned} \label{eq:3foldV}
V_u \ \hookrightarrow & \  \ \mathcal{V} \cr 
&\ \ \downarrow\cr 
& \ \  \mathbf{D}\,
\end{aligned}\end{equation}
of Calabi--Yau threefolds $V_u$ varying over the unit disk $\mathbf{D}=\{u\in\mathbb{C}\,|\,|u|\leq 1\}$, such that the total space $\mathcal{V}$ and the generic fibers $V_{u\neq 0}$ are smooth. The point $u=0$ models the infinite distance limit in the complex structure moduli space. The corresponding central fiber $V_0$ is not smooth, but rather decomposes into a union 
\begin{equation}
\label{eq:def-V0}
    V_0 = \bigcup_{i=1}^N V_i\,
\end{equation}
of $N$ irreducible \textit{smooth} components $V_i$ intersecting normally, for some positive integer $N$. The simple case $N=2$ is depicted in Figure \ref{fig:ss_degeneration}.

As described in \cite{Monnee:2025ynn}, the physical interpretation of the limit depends crucially on how the various components $V_i$ intersect each other. In the following, we will denote these intersections by
\begin{equation} \label{eq:Vi0ik}
    V_{i_0\cdots i_k} = V_{i_0}\cap\cdots \cap V_{i_k}\,,
\end{equation}
 and their union as 
 \begin{equation}\label{def:Vk+1-1}
    V^{(k+1)} = \bigsqcup_{i_0,\ldots,i_k} V_{i_0\cdots i_k}\,,\qquad 0\leq k\leq 3\,.
\end{equation}
These spaces are conveniently combined into the so-called \textit{dual (intersection) graph} $\Pi(V_0)$ of $V_0$. The latter is a simplicial complex that is defined inductively as follows. First, to each component $V_i$ one assigns a vertex. Then, for each connected component in the intersection $V_{ij}=V_i\cap V_j$ one assigns an edge connecting the respective vertices. Continuing on, one inductively assigns a $k$-simplex to each connected component of a non-empty intersection $V_{i_0\cdots i_k}$. In particular, if $d$ denotes the largest integer such that $V^{(d+1)}$ is non-empty, 
 the dual graph has dimension $d$. Importantly, whether the infinite distance limit under consideration is of type II, III, or IV in the nomenclature of \cite{Grimm:2018cpv} is simply determined by the values $d=1,2,3$, i.e.
\begin{equation}
    \boxed{\text{Type II / III / IV}\qquad \implies\qquad \text{$\Pi(V_0)$ has dimension 1/2/3\,.}}
\end{equation}
Let us stress that the physical interpretation of these three types of limits is very different. Indeed, as argued in \cite{Hassfeld:2025uoy,Monnee:2025ynn}, one finds
\begin{align}
    \begin{split}
    &\text{Type II}:\qquad \,\,\text{Emergent string limit}\,,\\
    &\text{Type III}:\qquad \text{Decompactification to 6d}\,,\\
    &\text{Type IV}:\qquad \text{Decompactification to 5d}\,.
    \end{split}
\end{align}
Let us also remark that at the level of geometry, the above discussion applies equally well, mutatis mutandis, to semi-stable degenerations of Calabi--Yau fourfolds. In that case, there is additionally the possibility of a type V limit, which corresponds to the dual graph having dimension 4. This will be relevant in Section \ref{ssec:Ftheorylift} when we discuss the F-theory lift of complex structure degenerations of Type IIB orientifolds. We refer to the companion paper~\cite{Paper2} for a more explicit description of semi-stable Calabi--Yau fourfold degenerations.

\subsection{Degenerations and orientifolds}\label{ssec:oriendegenerations}
Next, we discuss what happens to the aforementioned geometric interpretation of infinite distance limits after performing an orientifold projection with O7-planes. Importantly, the result will depend on how the orientifold acts on the generic fiber $V_u$ and, in particular, on the central fiber $V_0$. To characterise the latter, we will distinguish two different types of orientifolds, which we dub \textit{O-type A} and \textit{O-type B}. These differ by how the O7-planes are wrapped in the central fiber $V_0$ or, equivalently, by the orientifold action on the dual graph $\Pi(V_0)$ introduced in the previous section. As we will see, the distinction between O-type A and O-type B is enough to infer whether or not $g_s$ corrections become unsuppressed along the classical limit $\cM_{\rm c.s.}(V/\Omega)$.

\subsubsection{General discussion}
To characterise the two types of orientifolds, we specify the relative location of the orientifold fixed-point locus with respect to the intersection loci (\ref{eq:Vi0ik}) of the components of the degenerate threefold. The physics will depend on whether or not the divisor wrapped by the orientifold 7-plane has support along one of the intersection loci. Since all codimension-$k$ loci are contained in some of the loci of codimension $k' \leq k$, it suffices to specify the relative position of the O7-plane divisor and codimension-one intersection loci, i.e., the double surfaces $V_{i_0i_1}\subset V^{(2)}$. 
This motivates the following definition:
\begin{definition}[O-type A/B]\label{def:otypes}
    Consider Type IIB string theory compactified on a Calabi--Yau threefold $V$ that undergoes a semi-stable degeneration as in~\eqref{eq:def-V0}, and let us denote by $V^{(2)}$ the union \eqref{def:Vk+1-1} of double surfaces $V_{i_0i_1}$ of $V_0$.
    \noindent Modding out Type IIB string theory by the action of $\Omega_p(-1)^F \sigma$ yields an ${\rm O\text{-}type\,\,A\,\,orientifold}$ if the O7-plane locus, defined as the codimension-one stratum of the fixed point locus of $\sigma$, has support along at least one component of $V^{(2)}$. Otherwise, the orientifold is of ${\rm O\text{-}type\,\,B}$.
\end{definition}

\begin{figure}[t]
    \centering
    \begin{subfigure}{0.40\textwidth}
        \centering
        \includegraphics[width=0.83\textwidth]{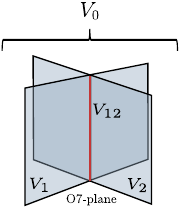}
        \caption{O-type A}
        \label{fig:O-type_A}
    \end{subfigure}
    \hfill
    \begin{subfigure}{0.40\textwidth}
        \centering
        \includegraphics[width=\textwidth]{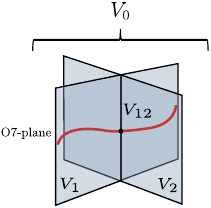}
        \caption{O-type B}
        \label{fig:O-type_B}
    \end{subfigure}
    \caption{An example of an O-type A/B orientifold in a two-component type II degeneration, in which the threefold splits into two components $V_1$ and $V_2$, intersecting over the double surface $V_{12}$. In the O-type A limit, the O7-plane (indicated in red) wraps the whole double surface $V_{12}$, whereas in the O-type B limit it intersects it in codimension 1.}
    \label{fig:O-types}
\end{figure}

For an illustration of the different O-types, see Figure \ref{fig:O-types}. To appreciate the meaning of the definition further, we recall that the dual graph of the degeneration $\Pi(V_0)$ is constructed by assigning a vertex to each component of $V_0$. The double surfaces $V_{i_0i_1}$ then correspond to the edges connecting the vertices. For the O7-plane to have support along a double surface $V_{i_0i_1}$, the geometric part, $\sigma$, of the orientifold action $\Omega=\Omega_p(-1)^F \sigma$ has to exchange the components $V_{i_0}$ and $V_{i_1}$. For this exchange to be a geometric symmetry, the dual graph $\Pi(V_0)$ must be symmetric under the exchange $V_{i_0}\leftrightarrow V_{i_1}$. Modding out by $\Omega$ then reduces the dual graph $\Pi(V_0)$ to its sublocus that is fixed under $\sigma$. For O-type A orientifolds, the O7-plane has support along at least one double surface $V_{i_0i_1}$ such that the dual graph of the degeneration is not point-wise invariant under $\sigma$. In particular, the number of components of $V_0$ is reduced by at least one. As a result, the fixed locus $\Pi(V_0)_{\rm fix}$ to which the orientifold projects the dual graph is not identical with $\Pi(V_0)$. By contrast, for O-type B limits the dual graph $\Pi(V_0)$ is point-wise invariant under the action of $\sigma$.

As a direct consequence of the relation between O-type A limits and the properties of $\Pi(V_0)$ under $\sigma$, all type IV limits have to be of O-type A after orientifolding. To see this, we can use the SYZ description of Calabi--Yau threefolds in the vicinity of type IV singularities as $T^3$-fibration over $S^3$~\cite{Strominger:1996it}. Whereas the original work~\cite{Strominger:1996it} conjectures that such a fibration structure holds in the vicinity of the large complex structure limit (where it can be argued for using mirror symmetry), it was shown to hold for general type IV degenerations in \cite{Monnee:2025ynn} without having to invoke mirror symmetry. Importantly, the $S^3$-base of the SYZ-fibration can be identified with the dual graph $\Pi(V_0)$ of the type IV degeneration. If we consider the orientifold action on the SYZ fibration, holomorphicity of the divisor supporting the O7-plane requires the O7-plane to have the same number of legs along base and fiber. For dimensional reasons this excludes the possibility of the O7-plane locus wrapping the entire $S^3$.  Since we identify the base $S^3$ with the dual graph of the degeneration, this means that $\Pi(V_0)$ cannot be point-wise invariant under the orientifold action such that, indeed, all type IV limits are of O-type A after orientifolding. 

Notice that for O-type A limits, the dimension of $\Pi(V_0)_{\rm fix}$ can be smaller than the dimension of $\Pi(V_0)$. This is a first signal that the primary singularity type of an O-type A limit can be reduced compared to the primary singularity type of the limit in the complex structure moduli space of the original Calabi--Yau threefold. This suggests that also the physical interpretation of the limit will be different in the orientifolded theory. This will be confirmed in a concrete example in Section \ref{ssec:Ftheorylift} by considering the lift to F-theory. 

Actually, to see that the physics of an O-type A limit -- and in fact also of an O-type B limit --  must change compared to original limit in the 4d $\cN=2$ parent theory one does not need to invoke F-theory. Already in the perturbative Type IIB orientifold picture we can consider the effect of the orientifold action on the three-cycles that shrink to zero size as $V$ degenerates. In~\cite{Hassfeld:2025uoy,Monnee:2025ynn} it was shown that in the vicinity of a component $V_{i_0\dots i_d}$ of the maximal codimension intersection $V^{(d+1)}$ inside $V_0$, the geometry is locally given by
\begin{equation}
    \cO(0)^{\oplus d} \hookrightarrow V_0^{\rm loc} \to V_{i_0\dots i_d}\,. 
\end{equation}
Using this, \cite{Hassfeld:2025uoy,Monnee:2025ynn} inferred the existence of special Lagrangian three-cycles $\Gamma$ inside $V$. Near the degeneration point in moduli space, these can be viewed as fibrations 
\begin{equation}\label{gammafib}
    T^d \hookrightarrow \Gamma \to B_{3-d}\,,
\end{equation}
where $B_{3-d}$ is a $(3-d)$-dimensional submanifold of the component $V_{i_0\dots i_d}$. Wrapping D3-branes on the cycles $\Gamma$ then produces BPS states that become massless in the infinite distance limit of the 4d $\cN=2$ theory. 
 The existence of this tower signals the emergence of a new duality frame, which is also reflected in the 
 asymptotic form of the couplings in the vector multiplet sector of the effective action.

The 3-cycles $\Gamma$ appearing in \eqref{gammafib} can be decomposed into an orientifold even and odd component, $\Gamma = \Gamma_+ + \Gamma_-$, where $\Gamma_{\pm} \in H^\pm_3(V)$.
In the orientifolded theory, the complex structure moduli which survive the projection are counted by the orientifold odd cohomology group $H^{2,1}_-(V)$.
  Since $\Gamma$ is a special Lagrangian 3-cycle, the surviving complex structure moduli therefore control the volume of $\Gamma_-$, while the volume of $\Gamma_+$ is invariant under changes of the orientifold odd moduli. 
  If $\Gamma_+ \neq 0$, then the volume of $\Gamma$ can vanish in the complex structure limit of the 4d ${\cal N}=1$ theory only if
  the orientifold projection fixes the moduli in $H^{2,1}_+(V)$ on a special locus where the volume of $\Gamma_+$ vanishes.\footnote{From the construction in \cite{Hassfeld:2025uoy,Monnee:2025ynn} one can convince oneself that this is only possible if $\Gamma$ does not have the topology of a 3-torus $T^3$.} 
  Furthermore, $\Gamma_+$ by itself cannot be a cycle whose volume vanishes only at infinite distance as otherwise we cannot deform $V$ away from the infinite distance degeneration after taking the orientifold projection -- a highly non-generic situation which we exclude.
  
The 4d particle state obtained from a D3-brane wrapped on $\Gamma = \Gamma_+ + \Gamma_-$ is mapped, by the orientifold action, to 
 a state that can be thought of as a bound state of a particle on $\Gamma_+$ and an anti-particle on $\Gamma_-$.
 In the orientifold theory, the potential tower of asymptotically massless states is therefore composed of particle-anti-particle bound states. 
 Apart from being unstable due to the appearance of a tachyon in the worldline quantum mechanics of the state, such objects are non-BPS and therefore their mass is no longer protected against quantum corrections.\footnote{To be precise, if $\Gamma_+ \neq 0$, the instability only concerns the component of the state from $\Gamma_-$. The resulting states from $\Gamma_+$ alone, however, cannot give rise to a weakly coupled gravitational tower because as explained above, the volume of $\Gamma_+$ vanishes at finite distance on $V$.}
This casts severe doubt on the appearance of an asymptotically massless tower in the orientifold theory and hence on the consistency of the limit.\footnote{Note that the projection of particle towers was already considered in \cite{EnriquezRojo:2020hzi}, but our conclusions differ.}
 
 To illustrate this point, consider a minimal type III limit of a Calabi--Yau threefold $V$, in the sense that the normal crossing variety $V_0$ splits into three components 
\begin{equation}
    V \to V_0 = V_1 \cup V_2 \cup V_3\,,\qquad V_{123} = V_1 \cap V_2 \cap V_3 \neq \emptyset\,.
\end{equation}
In this case $V^{(3)}$ has a single component $V_{123}$ that (by adjunction) is a genus-one curve and we can view the vicinity of $V_{123}$ as 
\begin{equation}\label{Vloc}
    \cO_{V_{12}}(0) \oplus \cO_{V_{23}}(0) \hookrightarrow V_0^{\rm loc} \to V_{123}\,.
\end{equation}
The first $\cO(0)$ fiber can be viewed as the normal direction of $V_{123}$ inside $V_{12}$ and the second as its normal direction inside $V_{23}$. There is a two-dimensional lattice of $T^3$s that shrink on $V_0$, constructed by fibering the $S^1$s in the $\cO(0)\simeq \mathbb{C}$ factors over the two-dimensional lattice of one-cyles $H_1(V_{123})$~\cite{Monnee:2025ynn}. Consider now an O-type A orientifold action exchanging $V_1\leftrightarrow V_2$ such that $V_{12}$ is the O7-plane locus. In the local description~\eqref{Vloc} of the degeneration, the orientifold wraps $V_{123}$ and $\cO_{V_{12}}(0)$. For this reason, the one-cycle in $H_1(V_{123})$ as well as the $S^1$ in $\cO_{V_{12}}(0)$ are invariant under the orientifold action. However, the $S^1$ in $\cO_{V_{23}}(0)$ is not invariant under the orientifold action. As a consequence, the entire two-dimensional lattice of $T^3$s is projected out by the orientifold action, and  so is the two-dimensional lattice of wrapped D3-brane states that are associated with the infinite distance limit in the 4d $\cN=2$ parent theory. These are the states whose masses vanish fastest in the infinite distance limit and at the rate characteristic of a type III limit; in the language of asymptotic Hodge theory, they are dual to elements in ${\rm Gr}_1$. The elimination of these states in the above sense
is a clear signal that the effective physics in the limit has to differ considerably from the 4d $\cN=2$ parent theory.

As we will find in the uplift to F-theory, the orientifold projection of the towers can manifest itself in two different ways: 
Either $\Gamma_-$ uplifts to a 4-cycle in the F-theory fourfold whose volume vanishes only in an infinite distance degeneration of the fourfold complex structure moduli, or not. In the latter case, the classical infinite distance limit is either manifestly absent in the full quantum theory including $g_s$ corrections, which are summed up in the fourfold geometry, or at least
 the severity of the degeneration changes drastically. For instance the primary type of degeneration can be reduced, as discussed, or the number of components can be smaller. These effects will be observed for O-type A orientifolds, but not for O-type B orientifolds.
Even if all leading $\Gamma_-$ uplift to suitable 4-cycles in the F-theory fourfold (as is the case for O-type B orientifolds), there is no candidate for an asymptotically light tower from the perspective of F/M-theory. Indeed, we will argue in \cite{Paper2} that even such situations are quantum obstructed, but by K\"ahler moduli effects (rather than $g_s$ effects), which are not visible in the F-theory geometry.

\subsubsection{Example: \texorpdfstring{$\mathbb{P}^4_{2,2,2,3,3}[12]/(\mathbb{Z}_6\times\mathbb{Z}_2^2)$}{P22233[12]/(Z6xZ2xZ2)}}
As an example, we discuss the mirror $V$ of the (resolved) degree 12 hypersurface Calabi--Yau in the weighted projective space $\mathbb{P}^4_{2,2,2,3,3}$, given by the zero locus
\begin{equation}
    \{P_V= x_1^6+x_2^6+ x_3^6 +x_4^4 +x_5^4 + \psi x_1x_2x_3x_4x_5 + \phi_1 (x_1x_2x_3)^2 + \phi_2 (x_4x_5)^2=0\}/G_{\rm GP}\,.
    \label{eq:P22233}
\end{equation}
Geometric details on this Calabi--Yau threefold can be found in Appendix~\ref{app:CYex}, where we compute the action of the associated Greene--Plesser group $G_{\rm GP}$ and argue that all degenerations studied in the following are semi-stable.\footnote{Notice that the complex structure deformations $\psi,\phi_1,\phi_2$ only span a subspace of the full complex structure moduli space of $\mathbb{P}^4_{2,2,2,3,3}[12]/(\mathbb{Z}_6\times\mathbb{Z}_2^2)$, which has total dimension 6. For the discussion in the following, the other non-polynomial complex structure deformations do not play a role.} In the following, we discuss the various infinite distance limits obtained by sending one of the parameters $\psi,\phi_1,\phi_2\to \infty$. We furthermore discuss possible orientifold actions under which these moduli are not projected out.

\subsubsection*{Type IV: $\psi\to\infty$}
In this limit, the threefold $V$ degenerates into a union of five smooth components,
\begin{equation}\label{eq:degquintic}
    V\to V_0^{\psi}=\bigcup_{i=1}^5 \{x_i=0\}\,.
\end{equation}
Before orientifolding, the dual graph of the degeneration characterising $V_0$ is a triangulation of $S^3$, namely a 5-cell. For the orientifold action, the simplest choices are either reflections $x_i\to -x_i$ or coordinate exchanges $x_i\leftrightarrow x_j$. In the simplest case, the orientifolds only lead to O7-planes and no O3-planes. This can be achieved either by reflecting a single coordinate or by exchanging a pair of coordinates. Given the defining polynomial~\eqref{eq:P22233}, we see that a reflection $x_i\to -x_i$ projects out the deformation $\psi$, producing an orientifold without complex structure deformations. Instead, if we exchange a pair of coordinates, the monomial multiplying $\psi$ is invariant. For definiteness, let us consider the geometric action of the orientifold to be the exchange $x_1\leftrightarrow x_2$. The O7-plane wraps the locus $\{x_1 = x_2\}$, which in particular includes the double surface $\{x_1=x_2=0\}$. The orientifold action is therefore of O-type A, in accordance with the observation that type IV limits only admit O-type A orientifolds.

The action of the orientifold on the dual graph exchanges two vertices of the 5-cell, corresponding to a folding of the latter to a solid tetrahedron. This tetrahedron is not a triangulation of an $S^3$, but rather of a 3-disk $\mathbf{D}^3$, showing that the orientifolded dual graph is still 3-dimensional, but with a reduced number of vertices. Such a geometry cannot arise as the dual graph of a degeneration of a Calabi--Yau threefold~\cite{Monnee:2025ynn,Kollar:2015}, signalling that the orientifolded limit is very different from the original one.

\subsubsection*{Type III: $\phi_1\to\infty$}
\begin{figure}[t]
    \centering
    \begin{subfigure}{0.45\textwidth}
        \centering
        \includegraphics[width=0.8\textwidth]{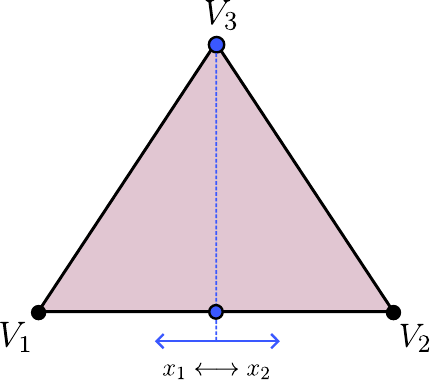}
        \caption{O-type A Type III}
        \label{fig:O-type_A-graphIII}
    \end{subfigure}
    \hfill
    \begin{subfigure}{0.45\textwidth}
        \centering
        \includegraphics[width=0.75\textwidth]{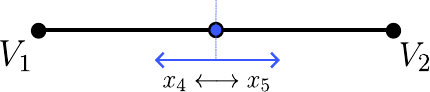}
        \caption{O-type A Type II}
        \label{fig:O-type_A-graphII}
    \end{subfigure}
    \caption{The dual graphs (a) $\Pi(V_0^{\phi_1})$ and (b) $\Pi(V_0^{\phi_2})$ corresponding to the type III limit $\phi_1\to\infty$ and the type II limit $\phi_2\to\infty$, respectively. The action of the two O-type A orientifolds and the resulting orientifolded dual graph are depicted in blue. In both cases the orientifold action reduces the dimension of the dual graph by one.}
    \label{fig:dual-graph_orientifold}
\end{figure}

In this limit, the Calabi--Yau threefold degenerates into the union of three smooth components,\footnote{As explained in Appendix~\ref{app:CYex}, there are two $\mathbb{Z}_3$ subgroups of $G_{\rm GP}$, generated by the transformations $(g^{(1)})^2$ and $((g^{(1)})^{-2}g^{(2)})^2$, whose action on $x_{1,2,3}$ reduces the three threefold components.}
\begin{equation}
    V\to V_0^{\phi_1}=\{x_1^2=0\}\cup \{x_2^2=0\}\cup \{x_3^2=0\}\,.
\end{equation}
These intersect pairwise over three surfaces, which in turn intersect over a single elliptic curve. The dual graph $\Pi(V_0^{\phi_1})$ is a solid triangle. An O-type A orientifold is obtained by swapping $x_i\leftrightarrow x_j$ for a pair $i,j\in\{1,2,3\}$. In this case, the O7-plane locus coincides for $\phi_1\to\infty$ with the intersection of $\{x_i^2=0\}$ and $\{x_j^2=0\}$.
Furthermore, the orientifold acts on the dual graph as indicated in Figure~\ref{fig:O-type_A-graphIII}, such that its fixed locus is an interval and has reduced dimension. By contrast, an O-type B orientifold is given by $x_4 \leftrightarrow x_5$. In the $\phi_1\to \infty$ limit this leaves all double surfaces invariant as a whole. In particular, the double surfaces are intersected by the O7-plane $\{x_4=x_5\}$ in codimension 1. We thus indeed have an O-type B type III limit. 

\subsubsection*{Type II: $\phi_2\to\infty$}
In this limit, the Calabi--Yau threefold degenerates into the union of two irreducible Fano components,
\begin{equation}
    V\to V_0^{\phi_2}=\{x_4^2=0\}\cup \{x_5^2=0\}\,,
\end{equation}
which intersect over a K3 surface. Notice that after modding out by the action of the Greene--Plesser group $G_{\rm GP}$, the two threefold components are reduced and hence smooth (see Appendix~\ref{app:CYex}). This shows that the dual graph $\Pi(V_0^{\phi_2})$ is an interval. An orientifold of O-type A that does not project out the modulus $\phi_2$ is obtained by the geometric action $x_4\leftrightarrow x_5$. In this case, the O7-plane locus coincides for $\phi_2\to\infty$ with the K3 surface that is the intersection of $V_1=\{x_4^2=0\}$ and $V_2=\{x_5^2=0\}$. In addition, since the orientifold swaps the components $V_1$ and $V_2$, it acts on the dual graph as indicated in Figure~\ref{fig:O-type_A-graphII}, such that the sublocus of the dual graph fixed under the involution is simply a point on the interval. Hence, the dimension of the dual graph is reduced by one. From the perspective of the dual graph, the orientifolded limit behaves as a type I limit, which is expected to be at {\it finite} distance.

Instead, an O-type B orientifold is obtained by considering a geometric action of the form $x_i\leftrightarrow x_j$ with $i,j\neq 4,5$. In this case, the O7-plane locus intersects the double K3 surface in codimension one. The full dual graph is point-wise invariant and the degeneration does not change due to the orientifold.

\subsection{Examples of F-theory lifts of O-Type A/B limits}\label{ssec:Ftheorylift}
So far, we have provided a geometric description of limits in the complex structure moduli space of a Calabi--Yau threefold and their interpretation after performing an orientifold. As described at the beginning of this section, the central question we would like to address in this work is whether such limits exist in the \textit{physical} quantum-corrected moduli space. A natural first step is to include $g_s$ corrections by considering the F-theory lift of the aforementioned Type IIB orientifolds. In this section, we construct this lift explicitly for the orientifolds of $\mathbb{P}^4_{2,2,2,3,3}[12]/(\mathbb{Z}_6\times\mathbb{Z}_2^2)$ discussed above. This example confirms the general discussion so far: For O-type A limits the number of components into which the F-theory fourfold splits is reduced compared to the Calabi--Yau threefold, while the lift is trivial for O-type B limits. A more detailed analysis including the physical interpretation of O-type A lifts will be undertaken in Section~\ref{sec:p-vs-np}.

As we have seen in Section \ref{ssec:oriendegenerations}, the Calabi--Yau threefold $V=\mathbb{P}^4_{2,2,2,3,3}[12]/G_{\rm GP}$ admits complex structure degenerations of types II, III, IV;
all three admit 
orientifolds of O-type A, while the first two also admit O-type B orientifolds. All orientifolds are given by exchanging two coordinates $x_i$, $x_j$ with the O-type depending on the complex structure degeneration. The first involution we consider, 
$x_i\leftrightarrow x_j$, $i,j\in\{1,2,3\}$, gives rise to an O-type A orientifold for the types III and IV limits and to O-type B for the type II limit; the second involution, 
exchanging $x_4\leftrightarrow x_5$, leads to O-type A for the type II and IV limits and to O-type B for type III limits. In the following, we lift both orientifolds to F-theory and study the degeneration of the F-theory base corresponding to the original limit. In Appendix~\ref{app:CYex} we show that all considered degenerations are semi-stable. 

\subsubsection{Involution 1: \texorpdfstring{$x_i\leftrightarrow x_j$, $i,j\in\{1,2,3\}$}{xi<->xj, i,j=1,2,3}}
Let us first consider the F-theory uplift for the orientifold action $x_i\leftrightarrow x_j$, $i\in\{1,2,3\}$, where for concreteness we focus on $\sigma:x_1\leftrightarrow x_2$. To start with, we introduce the change of variables
\begin{equation}
   v = x_1 + x_2 \,, \qquad w = x_1 - x_2  \,,
\end{equation}
and rewrite the defining hypersurface polynomial~\eqref{eq:P22233} as 
\begin{equation*}
    P_V = \frac{1}{32}\left(v^6+15v^4w^2+15v^2w^4+w^6\right)+x_3^6+x_4^4+x_5^4+\frac{\psi}{4}(v^2-w^2)x_3x_4x_5+\frac{\phi_1}{16}(v^2-w^2)^2x_3^2+\phi_2x_4^2x_5^2 \,.
\end{equation*}
The involution $\sigma$ now acts as $w\to-w$, with the O7-plane localised along the fixed divisor $D_w = \{w = 0\}$.

To construct the F-theory lift, one identifies 
the invariant monomials $(w^2,v,x_3,x_4,x_5)$ as new ambient space coordinates 
from which the base $\cB_3$ of an elliptic fourfold $W$ is cut out by a polynomial of suitable degree~\cite{Collinucci:2008pf,Collinucci:2008zs,Blumenhagen:2009up, Collinucci:2009uh}.
 With
 \begin{equation}
     h := w^2\,,
 \end{equation}
 the toric ambient space of $\cB_3$ is therefore
 $\mathbb{P}^4_{4,2,2,3,3}/G^\sigma_{\rm GP}$ with homogeneous coordinates
  $[h : v : x_3 : x_4 : x_5]$. Here, $G^\sigma_{\rm GP}$ denotes the subgroup of elements in $G_{\rm GP}$ that are invariant under the exchange $\sigma:x_1\leftrightarrow x_2$, i.e., here $G^\sigma_{\rm GP}=\mathbb{Z}_2^2$. In this ambient space, $\cB_3$ is the hypersurface with defining equation
\begin{equation}
    P_{\cB_3} = \frac{1}{32}\left(v^6+15v^4h+15v^2h^2+h^3\right)+x_3^6+x_4^4+x_5^4+\frac{\Tilde{\psi}}{4}(v^2-h)x_3x_4x_5+\frac{\Tilde{\phi_1}}{16}(v^2-h)^2x_3^2+\Tilde{\phi_2}x_4^2x_5^2 \,.
    \label{eq:P42233-12}
\end{equation}
Its anti-canonical class is given by
\begin{equation}
    {\bar K}_{\cB_3} = \frac{1}{2} [\{h=0\}] \,,
\end{equation}
which by adjunction is of degree $2$ for this orientifold action. 

The elliptically fibered Calabi--Yau fourfold $W$ over the base $\cB_3$ has an associated Weierstrass model which is given by the complete intersection
\begin{equation}
    W=\left(\{P_W=0\}\cap\{P_{\cB_3}=0\}\right)/G^\sigma_{\rm GP}
    \label{eq:fourfold}
\end{equation}
inside a $\mathbb{P}^2_{2,3,1}$-fibration over $\mathbb{P}^4_{4,2,2,3,3}/G^\sigma_{\rm GP}$. Here, $P_W=y^2-(x^3+fxz^4+gz^6)$ is the Weierstrass polynomial, defined in terms of homogeneous coordinates $[x:y:z]$ of the $\mathbb{P}^2_{2,3,1}$-fiber and sections $f$ and $g$ of ${\bar K}_{\cB_3}^4$ and ${\bar K}_{\cB_3}^6$, respectively. The elliptic fiber degenerates over the vanishing locus of the discriminant 
\begin{equation}
\Delta = 4 f^3 + 27 g^2\,. 
\end{equation}
The specific form of the functions $f$ and $g$ depends on the cancellation of the 7-brane tadpole. We will mostly focus on orientifold models in which this tadpole is cancelled globally, but not locally, by a single D7-brane along a divisor in the class $8[D_w]$ on $V$. The associated Weierstrass is generic, i.e., $f$ and $g$ are generic polynomials of degree $4\deg[\bar{K}_{\cB_3}]=8$ and $6\deg[\bar{K}_{\cB_3}]=12$ invariant under $G^\sigma_{\rm GP}$, respectively. In this case, the fourfold $W$ is smooth: Over generic points of $\Delta$ the elliptic fiber acquires an I$_1$ Kodaira singularity and is smooth away from $\Delta$. 
 We will discuss the associated Weierstrass models in more depth in Section \ref{sec:OTypeAEx}.

Before turning to the explicit degenerations, it is important to stress that while the relevant monomials in the defining polynomials of $V$ and $\cB_3$ look similar, the complex structure moduli $\psi,\phi_{1,2}$ of $V$ are {\it different} from the complex structure moduli $\Tilde{\psi},\Tilde{\phi}_{1,2}$ of the F-theory base $\cB_3$. This means in particular that the limits $\psi\to\infty$ and $\Tilde{\psi}\to\infty$ (and similarly for $\phi_{1,2}$) are two {\it different} limits in the F-theory moduli space. After all, the original Calabi--Yau threefold $V$ exists only in a very specific corner of the F-theory moduli space. We will expand on this point in Section~\ref{sec:p-vs-np}.

\paragraph{Type II O-type B limit.} Taking the type II limit $\Tilde{\phi}_2\to\infty$, we find that the two (reduced, after modding out $G^\sigma_{\rm GP}$) threefolds $V_i=\{x_i^2=0\}$, $i=4,5$, into which the F-theory base $\cB_3$ splits are invariant under the orientifold action.\footnote{Notice that the two components $V_i$ intersect the singular locus of the ambient weighted projective space. As we argue in Appendix~\ref{app:CYex}, the conclusions drawn in this subsection are unchanged when we resolve these ambient space singularities and then consider the proper transform of the singular fiber of the degeneration.} Correspondingly, the fourfold $W$ splits into two fourfold components $\Hat{V}_i$, $i=4,5$, obtained by restricting the elliptic fibration to the threefold components $V_i$. As there is a single double threefold $\Hat{V}_{45}=\Hat{V}_4\cap \Hat{V}_5$, the dual graph of the degeneration is an interval and, as anticipated from the general discussion above, this O-type B limit lifts trivially to F-theory.

\paragraph{Type III O-type A limit.} Taking the limit $\tilde{\phi}_1\to \infty$ in the complex structure moduli space of the F-theory fourfold $W$, it is evident that the base $\cB_3$ splits into \textit{two} components\footnote{The action of $g^{(2)}$, $g^{(3)}$ remains diagonal after changing coordinates to $[h:v:x_3:x_4:x_5]$ so that the two components $V_{v^2-h}$ and $V_3$ are reduced and hence smooth after modding by $G^\sigma_{\rm GP}$.} $V_{v^2-h}=\{(v^2-h)^2=0\}$ and $V_3=\{x_3^2=0\}$. Again we find that also the fourfold $W$ splits into two components $\Hat{V}_{v^2-h}$, $\Hat{V}_3$. Thus, from the fourfold perspective, the limit $\Tilde{\phi}_1\to\infty$ is a type II degeneration.

\paragraph{Type IV O-type A limit.} In the limit $\tilde{\psi}\to\infty$ the base $\cB_3$ splits into the four components $V_{v^2-h}=\{v^2-h=0\}$, $V_i=\{x_i=0\}$, $i=3,4,5$, so that the fourfold $W$ degenerates into the union of four \textit{smooth} elliptic fourfolds,
\begin{equation}
    W\longrightarrow\Hat{V}_{v^2-h}\cup\Hat{V}_3\cup\Hat{V}_4\cup\Hat{V}_5,
\end{equation}
where each fourfold component is obtained by restricting the elliptic fibration to the corresponding threefold component of $\cB_3$. The fourfold components intersect pairwise over six smooth double threefolds, which in turn intersect over $4$ smooth triple surfaces. These surfaces intersect over one copy of the elliptic fiber of $W$, which is localised over the intersection point
\begin{equation}
    V_{v^2-h}\cap V_3\cap V_4\cap V_5\subset\cB_3 
\end{equation}
on the base. As this point does not lie on $\{h=0\}$, we conclude that all intersection loci of the fourfold components are smooth and the limit $\tilde{\psi}\to\infty$ describes a good type IV degeneration of the fourfold $W$. While for the O-type A type III limit $\tilde{\phi}_1\to\infty$, the primary type of the singularity was reduced, we see  here that the primary type remains IV also from the fourfold perspective. Nevertheless, the limit has changed because the central fiber of the degeneration now splits into 4 instead of 5 components.

\subsubsection{Involution 2: \texorpdfstring{$x_4\leftrightarrow x_5$}{x4<->x5}}
Next, we construct the F-theory uplift for the orientifold action $x_4\leftrightarrow x_5$. Again we change the variables to
\begin{equation}
    p=x_4+x_5\,,\qquad q=x_4-x_5,
\end{equation}
in which the defining polynomial takes the form
\begin{equation}
    P=x_1^6+x_2^6+x_3^6+\frac{1}{8}\left(p^4+6p^2q^2+q^4\right)+\frac{\psi}{4}x_1x_2x_3(p^2-q^2)+\phi_1x_1^2x_2^2x_3^2+\frac{\phi_2}{16}(p^2-q^2)^2\,.
\end{equation}
The invariant monomials $(x_1,x_2,x_3,p,k=q^2)$ serve as coordinates on the space $\mathbb{P}^4_{2,2,2,3,6}/G^\sigma_{\rm GP}$ in which the F-theory base $\cB_3$ is the hypersurface cut out by
\begin{equation}
    P_{\cB_3}=x_1^6+x_2^6+x_3^6+\frac{1}{8}\left(p^4+6p^2k+k^2\right)+\frac{\Tilde{\psi}}{4}x_1x_2x_3(p^2-k)+\Tilde{\phi}_1 x_1^2x_2^2x_3^2+\frac{\Tilde{\phi}_2}{16}(p^2-k)^2\,,
    \label{eq:P22233-45}
\end{equation}
and
\begin{equation}\label{eq:Ofold2-Kbar}
    [\Bar{K}_{\cB_3}]= \frac{1}{2}[\{k=0\}]\,,
\end{equation}
which here is of degree $3$. The F-theory fourfold is again given by \eqref{eq:fourfold}. Notice that when expressed in terms of the generators $g^{(1)}$, $\Tilde{g}^{(2)}$ and $\Tilde{g}^{(3)}$ given in~\eqref{eq:P22233-45-GGP}, the full Greene--Plesser group is invariant under $\sigma$, $G^\sigma_{\rm GP}=G_{\rm GP}$.

\paragraph{Type II O-type A limit.} In this second orientifold, the limit $\tilde{\phi}_2\to\infty$ describes an O-type A limit. From \eqref{eq:P22233-45} we see that the base $\cB_3$ degenerates into $\{(p^2-k)^2=0\}$. Due to the fact that $G^\sigma_{\rm GP}=G_{\rm GP}$ and the chosen generators remain diagonal and non-trivial in the F-theory coordinates $[x_1:x_2:x_3:p:k]$, we find that the base $\cB_3$, and hence the F-theory fourfold $W$, does not split at all at the degeneration $\tilde{\phi}_2\to\infty$. In other words, after lifting the limit from the Type IIB orientifold to F-theory it is now of type I and correspondingly describes a {\it finite} distance singularity in the complex structure moduli space of $W$.

\paragraph{Type III O-type B limit.} Similar to the O-type B limit studied for the $x_1\leftrightarrow x_2$ orientifold above, in the limit $\tilde{\phi}_1\to\infty$ the base $\cB_3$ degenerates into three components $V_i=\{x_i^2=0\}$, $i=1,2,3$, which are invariant under the orientifold projection $x_4\leftrightarrow x_5$. Therefore, the limit lifts trivially to F-theory.

\paragraph{Type IV O-type A limit.} Comparing the defining polynomial \eqref{eq:P22233-45} in the current orientifold with the one obtained for the $x_1\leftrightarrow x_2$ orientifold in~\eqref{eq:P42233-12}, we see that the terms containing the complex structure modulus $\tilde{\psi}$ are identical and hence lead to the same type IV limit of the Calabi--Yau fourfold $W$. Let us stress again that even though the limit continues to be of type IV, the limit in the fourfold moduli space is different from the $\psi\to \infty$ limit in the moduli space of the original Calabi--Yau threefold as there are now only 4 instead of 5 components of the singular fiber.

\section{The \texorpdfstring{$g_s$}{gs}-corrected moduli space}\label{sec:p-vs-np}
In the previous section we have described how an infinite distance limit in $\cM_{\rm c.s.}(V/\Omega)$ uplifts to F-theory. Geometrically, we have found that O-type B limits uplift trivially to F-theory in the sense that the semi-stable degeneration of the Calabi--Yau threefold underlying the Type IIB orientifold remains unchanged in the F-theory lift. In particular, the elliptic fiber of the Calabi--Yau fourfold is merely a spectator in the F-theory lift of O-type B limits. By contrast, for O-type A limits the properties of the degenerate Calabi--Yau fourfold differ significantly from the semi-stable degeneration of the Type IIB Calabi--Yau threefold. Most importantly, the number of components into which the Calabi--Yau fourfold splits is smaller than the number of components of the original degenerate Calabi--Yau threefold. In some cases this can even lead to a change in the primary singularity type. This is an indication that for O-type A limits the perturbative Type IIB orientifold description does not correctly capture the physics. 

The goal of this section is to determine the physical interpretation of the F-theory lift of O-type A limits of Calabi--Yau orientifolds. Since F-theory is a non-perturbative description of Type IIB string theory, the expectation is that the F-theory lift of O-type A limits encodes effects that go beyond the perturbative Type IIB description. The aim of the following analysis is to work out the difference between the perturbative Type IIB description and its non-perturbative completion within F-theory to delineate the regime of applicability of a perturbative Type IIB description in the 4d $\cN=1$ moduli space. 
\subsection{General analysis}\label{ssec:general}

We first recall how the perturbative Type IIB moduli space is embedded into the complex structure moduli space of the F-theory Calabi--Yau fourfold. For simplicity, we focus on a two-dimensional subspace of two moduli spaces.  
 To this end, we single out one complex structure modulus $z_B=-\frac{1}{2\pi \ii} \log \phi$; the subscript indicates that this is a modulus of the orientifold in the perturbative Type IIB description of the theory. The modulus $z_B$ is inherited from the Calabi--Yau threefold $V$ whose orientifold is taken. In the limit $z_B\to \ii \infty$, the Calabi--Yau threefold $V$ degenerates as 
\begin{equation}\label{eq:degV}
    V\longrightarrow V_0=\bigcup_{i=1}^N V_i\,,
\end{equation}
which we take to be a type $\mathtt{A}$ limit where $\mathtt{A}\in\{{\rm II,\,III,\,IV}\}$. In the examples discussed above, the degenerate Calabi--Yau threefold corresponds to a hypersurface of the form
\begin{equation}
    V_0 =\{\phi \prod_{j} x_j^{k_j}=0\} \,. 
\end{equation}
The subspace of the Type IIB moduli space that we focus on is then parametrised by the 
\begin{equation}
    {\rm perturbative\;Type\;IIB \; moduli}: \qquad \left(z_B=\frac{\ii}{2\pi} \log \phi, \,  \tau=C_0+\frac{\ii}{g_s}\right)\,,
\end{equation}
where $\tau$ is the Type IIB axio-dilaton. Now consider the lift of this Type IIB orientifold setup  to F-theory compactified on the elliptically fibered Calabi--Yau fourfold 
\begin{equation}
    \mathcal{E}\hookrightarrow W\to \cB_3\,,
\end{equation}
where $\cE$ denotes the elliptic fiber, and $\cB_3$ denotes the base threefold. In the F-theory lift, there will be two relevant complex structure moduli of the Calabi--Yau fourfold, which we denote by the
\begin{equation}
    \text{F-theory\; moduli:}\qquad \left(z_F,z_S\right)\,.
\end{equation}
Here $z_F$ is the F-theory lift of the modulus $z_B$ and describes the limit in the complex structure moduli space of the base $\cB_{3}$ that we associate classically with $\phi\to \infty$. As we saw in previous sections, the main difference between O-type A and B limits is the asymptotic form of the base $\cB_{3,0}$\footnote{More generally, for O-type A limits the prefactor could be $(p^2-k)^{n}$ for some power $n\geq1$. After modding out the orientifold-invariant Greene--Plesser group $G^\sigma_{\rm GP}$, this factor is reduced. For the following analysis we work with the reduced component.}
\begin{align}
 &     \text{O-type A:}\qquad \tilde{\phi} (p^2-k)^2 \prod_{j'}  x_{j'}^{k_{j'}} =0 \,,\label{degenerationOA} \\
& \text{O-type B:}\qquad \tilde{\phi} \prod_{j} x_j^{k_j}=0\,.\label{degenerationOB}
 \end{align}
 Here $j'$ runs over the ambient space coordinates different from $p$ and $k$; $k_j\geq0$ and $\tilde{\phi}$ correspond to the complex structure parameter that is taken to infinity. In the O-type B case at least two of the $k_j$ are non-zero whereas for O-type A limits all $k_j$ can vanish. The modulus $z_F$ is then given by 
\begin{equation}
    z_F = \frac{\ii}{2\pi} \log \tilde\phi \,.
\end{equation}
In addition, the modulus $z_S$ is associated with the Sen-limit, which corresponds to $z_S\to \ii\infty$. In this limit the perturbative Type IIB picture is recovered -- at least for finite $z_F$. In the following, we describe the Sen-limit as a semi-stable degeneration limit of the fourfold $W$. As discussed in~\cite{Clingher:2012rg}, the elliptic fiber $\cE$ degenerates (after blow-up) over a generic point in $\cB_3$ into the union of two rational curves intersecting over two points. Hence, the total fourfold degenerates as
\begin{equation}\label{eq:Sen}
    W\longrightarrow W_0=W_1\cup W_2\,,
\end{equation}
where each $W_{i}$ is a $\mathbb{P}^1$-fibration over $\cB_3$. Importantly, the intersection of the two fourfold components, 
\begin{equation}
    V:=W_1\cap W_2\,,
\end{equation}
is itself a Calabi--Yau threefold, and corresponds to the double cover of the base $\cB_3$, such that we can identify $\cB_3=V/\mathbb{Z}_2$ as the Type IIB orientifold. Furthermore, the branch locus of the double cover corresponds to the O7-plane locus in $V$. Instead of $z_S$ we can also use the local coordinate $\epsilon$, which is related to $z_S$ via
\begin{equation}
    j(z_S)\sim \frac{1}{\epsilon^2} \,,\quad\rightarrow\quad  z_S = \frac{1}{\pi \ii} \log \epsilon+\mathcal{O}(\epsilon) \,.
\end{equation}
Here $j$ is the $SL(2,\mathbb{Z})$ invariant $j$-function. Classically, we identify $z_S$ with $\tau$ and $z_F$ with $z_B$. 

Let us denote by ${\Delta}_S=\{\epsilon=0\}$ and ${\Delta}_F=\{\tilde{\phi}=\infty\}$ the divisors in the complex structure moduli space of $W$ in which either the Sen-limit is taken or the base degenerates as in \eqref{degenerationOA} or \eqref{degenerationOB}. If the tree-level Type IIB effective action described the theory in the vicinity of $\Delta_F$ and $\Delta_S$ correctly, these two divisors would intersect normally; this would be a manifestation that at tree-level in the perturbative Type IIB effective action, the moduli space factorises into components describing the Type IIB axio-dilaton, the complex structure deformations of $V/\Omega$ and the open string moduli. Depending on the O-type of the limit, this factorisation of the moduli space might or might not hold at the quantum level. In terms of the divisors $\Delta_S$ and $\Delta_F$, a lack of factorisation is reflected in a point of tangency between these divisors.

The key feature of O-type A limits is that the divisor supporting the O7-plane becomes non-reduced. This is easiest understood in the Sen-limit of the F-theory uplift, where the location of the O7-plane is given by the divisor 
$k=0$, with $k$ the variable appearing in (\ref{degenerationOA}).
  In the O-type A limit, as $\tilde \phi \to \infty$, we set $k=p^2$, and hence the single divisor $k=0$ splits into a stack of two coincident divisors in class $p=0$.
 See also the example in Section \ref{sec:OTypeAEx} for more details of this general phenomenon.
In O-type B limits such a factorisation occurs only over a curve.  In the following, we focus on the more interesting O-type A case.

 The factorisation of the O7-plane locus 
 signals a non-perturbative deformation to a special locus in the 7-brane moduli space. 
A limit in the F-theory moduli space in which the base $\cB_3$ factorises as in~\eqref{degenerationOA} therefore corresponds, from the Type IIB perspective, to a limit in the open string moduli space or to a combined limit for open and closed string moduli.
 To reflect this lack of moduli space factorisation, the divisors $\Delta_S$ and $\Delta_F$ must intersect in a point of tangency. Notice that the limit 
 can lead to further singularities in $W$ but, depending on the concrete choice of Weierstrass model, does not have to.  This is a further indication that the modulus responsible for the degeneration~\eqref{degenerationOA} encodes open-string information.\footnote{Let us stress that the above reasoning only applies to O-type A limits, where the factorisation of $\{k=0\}$ occurs in codimension-one.  In O-type B limits, $\{k=0\}$ factorises in higher codimension, and the divisors $\Delta_F$ and $\Delta_S$ hence intersect normally, as expected classically.}

\begin{figure}[t]
    \centering
    \begin{subfigure}{0.20\textwidth}
        \centering
        \raisebox{0.33cm}{
        \includegraphics[width=\textwidth]{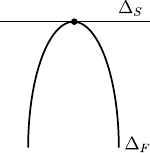}}
        \caption{}
        \label{subfig:tangency}
    \end{subfigure}
    \hspace{4cm}
    \begin{subfigure}{0.30\textwidth}
        \centering
        \includegraphics[width=\textwidth]{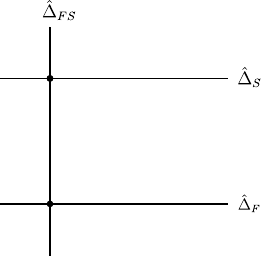}
        \caption{}
        \label{subfig:tangency-resolution}
    \end{subfigure}
    \caption{(a) For O-type A limits, the divisors $\Delta_S$ and $\Delta_F$ intersect in a point of tangency from the Type IIB perspective. (b) In the F-theory moduli space, the point of tangency can be resolved via blow-up, introducing an exceptional divisor $\hat{\Delta}_{FS}$.}
    \label{fig:tangency+resolution}
\end{figure}

The moduli space with the point of tangency between $\Delta_S$ and $\Delta_F$ is illustrated in Figure~\ref{subfig:tangency}. A point of tangency between the divisors $\Delta_S$ and $\Delta_F$ in the moduli space signals that the theories associated with these two divisors do not have compatible perturbative expansions. Indeed, at generic points along $\Delta_S$, the theory is well-described by the perturbative Type IIB effective action with vanishing string coupling and an O7-plane along $\{k=0\}$. If in addition to the Sen-limit we take the $\tilde{\phi}\to \infty$ limit, the O7-plane locus factorises. This implies that locally, a stack of two O7-planes coincides. However, two coincident O7-planes are highly non-perturbative from the Type IIB perspective. Thus, the perturbative Type IIB effective action that describes the theory at generic points along $\Delta_S$ is invalid at the point of tangency $\Delta_S\cap \Delta_F$ due to the non-perturbative effects associated with coincident O7-planes. 

A consistent perturbative description of the intersection $\Delta_S \cap \Delta_F$ requires normal crossing divisors in the moduli space. This is achieved by a blow-up of the point of tangency.\footnote{Such resolutions of points of tangency in the complex structure moduli space are well-studied in Calabi--Yau threefolds. A classic example is the resolution of the point of tangency between the discriminant and a type II divisor in the mirror of $\mathbb{P}^4_{1,1,2,2,6}[12]$, first discussed in~\cite{Kachru:1995fv} and more recently in~\cite{Lee:2019oct,Monnee:2025msf,Castellano:2026bnx,Hattab:2026lho}.} The resolution removes the point of tangency from the moduli space and replaces it with an exceptional divisor which we denote by $\hat{\Delta}_{FS}$, see Figure~\ref{subfig:tangency-resolution}.\footnote{There might be additional exceptional divisors appearing in the resolution process, but for us only one of these divisors is relevant.} In the resolution process, the divisors $\Delta_F$ and $\Delta_S$ transform into the divisors $\hat{\Delta}_F$ and $\hat{\Delta}_S$ in such a way that the divisor $\hat \Delta_S$ agrees with the divisor $\Delta_S$ outside the neighbourhood of the intersection $\hat\Delta_{S}\cap \hat{\Delta}_{FS}$, and similarly for $\hat \Delta_F$. This means that the theories corresponding to points along $\hat{\Delta}_S$ away from the intersection with $\hat{\Delta}_{FS}$ can be identified with theories along $\Delta_S$, i.e., are described by the perturbative Type IIB effective action in the $g_s\to 0$ limit. By contrast, the theories corresponding to points in the vicinity of the intersection $\hat{\Delta}_{FS}\cap \hat{\Delta}_S$ cannot be described within perturbative Type IIB string theory. While there does exist a perturbative description for these theories, the expansion parameter for the perturbation theory is not the classical string coupling $g_s$. 

The naive expectation from the Type IIB perspective would be that the intersection $\Delta_F\cap \Delta_S$ is of type $\mathtt{A}+{\rm I}$, reflecting the enhancement of the singularity from type from $\mathtt{A}$ along
$\Delta_F$ by collision with the type II singularity on $\Delta_S$.
However, since this point is removed in the blow-up procedure and replaced by the divisor $\hat{\Delta}_{FS}$, the intersection $\hat\Delta_{S}\cap \hat{\Delta}_{FS}$ does not necessarily have to be of type $\mathtt{A}+{\rm I}$. Below, we give an example of a model where this is indeed not the case. In fact, in this example the naive infinite distance degeneration of the Type IIB Calabi--Yau threefold is completely absent in the F-theory moduli space, even at $\hat \Delta_S \cap \hat \Delta_{FS}$. Our interpretation of this phenomenon is that, as the O7-plane begins to factorise into a stack of two coincident O7-planes in the limit, the model becomes so non-perturbative in the vicinity of the O7-plane that it is impossible to set $g_s =0$ by imposing the Sen-limit while at the same time taking the factorisation limit. In the example, the result is that 
the primary singularity type at $\hat\Delta_{S}\cap \hat{\Delta}_{FS}$ continues to be of type II (from the Sen-limit) and the naive factorisation limit of Type IIB theory is not part of the quantum corrected moduli space.

This phenomenon can also be cast in the language of period vectors: From the perturbative Type IIB perspective, we expect that the period vector of the fourfold $W$ in the closed string sector takes the form 
\begin{equation}\label{eq:periodsclass}
    \Pi_W = \begin{pmatrix} 1 \\\tau\end{pmatrix} \otimes \Pi_V \,, 
\end{equation}
where $\Pi_V$ are the periods of the Calabi--Yau threefold underlying the Type IIB orientifold. 
 In the vicinity of $\hat{\Delta}_S$, but away from the locus $\hat\Delta_{FS}$,
 this is indeed the case.
 However, after the resolution of the moduli space, the periods of $W$ in the vicinity of the locus $\hat\Delta_{S}\cap \hat{\Delta}_{FS}$ do not necessarily take the simple form~\eqref{eq:periodsclass} since the theory at this locus cannot be described using the perturbative Type IIB effective action. 
 Interestingly, this result is independent of the way how the tadpole is cancelled (locally by D7-branes on top of the O7-plane versus globally). In this respect our analysis seems to differ from \cite{Kim:2022jvv}, which concludes that if all tadpoles are cancelled locally, the factorisation \eqref{eq:periodsclass} persists throughout moduli space.

To summarise, the $g_s$-corrected moduli space of the perturbative Type IIB effective action is given by the complex structure moduli space of the fourfold $W$. This moduli space has tangencies between the Sen-limit divisor and the uplift of infinite distance divisors in $\cM_{\rm c.s.}(V/\Omega)$ corresponding to O-type A limits. This reflects the fact that the factorisation of the perturbative Type IIB moduli space into the complex structure moduli of $V$, the axio-dilaton $\tau$, and open string modulus does not prevail at the quantum level. In the F-theory moduli space, the tangencies can be resolved to obtain normal crossing divisors. This resolution is, however, not describable with the perturbative Type IIB effective action. As a result, the theory in the vicinity of the exceptional divisor in the F-theory moduli space cannot be understood using the perturbative Type IIB effective action.\footnote{This is similar in spirit to the resolution in the complex structure moduli space of the mirror of $\mathbb{P}^4_{1,1,2,2,6}[12]$ discussed in~\cite{Kachru:1995wm}. As explained in~\cite{Monnee:2025msf}, this resolution is also not describable using the perturbative heterotic dual as it involves the smooth nucleation of spacetime-filling NS5-branes.} In particular, the factorisation of the moduli space between the axio-dilaton, the closed string and the open string moduli is badly broken in the vicinity of $\hat{\Delta}_{FS}$ and $\hat{\Delta}_F$. 

From the perturbative type IIB perspective, we would have identified the divisor $\hat{\Delta}_F$ with the infinite distance divisor of $\cM_{\rm c.s.}(V/\Omega)$. However, as we argued, for O-type A limits the divisor $\hat{\Delta}_F$ is not accessible within the region of validity of the perturbative theory. The O-type A limit in $\cM_{\rm c.s.}(V/\Omega)$ is hence $g_s$-obstructed in the sense of Definition~\ref{def:2fusionheunconnected}. Moreover, even the locus $\hat{\Delta}_{FS}\cap \hat{\Delta}_S$ is not accessible within perturbative Type IIB string theory. Thus, the second condition in Definition~\ref{def:2fusionheunconnected} can actually not be realized at all in an O-type A limit.

\subsection{Example: O-Type A Type II limit on \texorpdfstring{$\mathbb{P}^4_{2,2,2,3,3}[12]/(\mathbb{Z}_6\times\mathbb{Z}_2^2)$}{P22233[12]/(Z6xZ2xZ2)}} \label{sec:OTypeAEx}
To illustrate the above general discussion in a concrete setup, we now revisit the model based on an orientifold of the mirror of $\mathbb{P}_{2,2,2,3,3}^4[12]$ presented in Section \ref{ssec:Ftheorylift}. For definiteness, we only focus on Orientifold 2, $x_4\leftrightarrow x_5$, and on the type II limit corresponding to $\phi_2\to \infty$, which yields an O-type A type II limit. As described in Section \ref{ssec:Ftheorylift}, the original type II O-type A limit $\phi_2\to\infty$ for the Calabi--Yau threefold lifts to a type I limit $\tilde{\phi}_2\to\infty$ for the F-theory base, which lies at finite distance. We will first show that, generically, this limit is in fact of type I$_{0,0}$ in the notation of~\cite{Grimm:2019ixq}, i.e., the fourfold $W$ does not degenerate at all along generic points of the locus $\hat \Delta_F$ in the F-theory complex structure moduli space. Furthermore we will analyse the behaviour at the intersections 
$\hat \Delta_F \cap \hat \Delta_{FS}$ and $\hat \Delta_S \cap \hat \Delta_{FS}$. We will conclude that there is no point in the $g_s$ quantum corrected moduli space in which the naive Type IIB picture of a factorisation of the Calabi--Yau threefold can be maintained.

\begin{figure}[t!]
    \centering
    \includegraphics[width=0.9\linewidth]{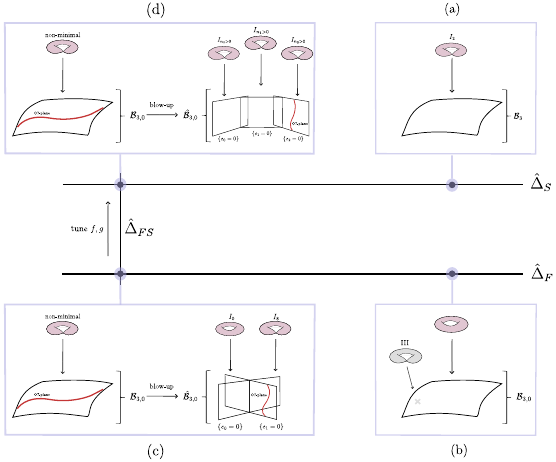}
    \caption{An illustration of the quantum moduli space for the F-theory lift of the type II O-type A limit $\phi_2\to\infty$ in the mirror of $\mathbb{P}^4_{2,2,2,3,3}[12]$. (a) A generic point on $\hat{\Delta}_S$, where the fourfold $W$ undergoes the standard Sen-limit. (b) A generic point on $\hat{\Delta}_F$, where the base becomes $\mathcal{B}_{3,0}=\{(p^2-h)^2=0\}/G_{\mathrm{GP}}$, which is smooth. For generic $g$, $W$ is smooth, but can acquire Kodaira type III singularities in codimension-3 (indicated in grey) for non-generic $g$. (c) The non-minimal Kodaira fiber and its resolution at $\hat{\Delta}_F\cap \hat{\Delta}_{FS}$. (d) The non-minimal Kodaira fiber and its resolution at $\hat{\Delta}_S\cap \hat{\Delta}_{FS}$. In (c) and (d) blowing down the component $\{e_0 =0\}$ leads to a semi-stable model.}
    \label{fig:p-vs-np_example}
\end{figure}

As reviewed in Section \ref{ssec:Ftheorylift}, the F-theory uplift of the orientifold  is defined in terms of a Weierstrass model over the base $\cB_3$.  The ambient space of $\cB_3$ is $\mathbb{P}^4_{2,2,2,3,6}/G^\sigma_{\rm GP}$ with homogeneous coordinates $[x_1:x_2:x_3:p:k]$, where for this choice of orientifold $G^\sigma_{\rm GP}=G_{\rm GP}=\mathbb{Z}_6\times\mathbb{Z}_2^2$. 
Let us now analyse the otherwise maximally generic Weierstrass model over the various enhancement loci in complex structure moduli space. The geometries arising at these loci are illustrated in Figure~\ref{fig:p-vs-np_example}.
\paragraph{Behaviour along $\hat \Delta_{F}$.} 
 We begin with the limit $\tilde{\phi}_2\to\infty$ in the F-theory moduli space with all other moduli
 at generic values and, in particular, away from weak coupling.
 To this end, we parametrise the base hypersurface as 
\begin{equation} \label{basedef}
    \cB_{3,\tilde \phi_2} =\{\tilde \phi_2(k-p^2)^2 + Q_{12}=0\}\subset \mathbb{P}^4_{2,2,2,3,6}/G_{\rm GP}\,,
\qquad Q_{12}\, \,  \text{generic}\,.
\end{equation}
Notice that $(k-p^2)^2$ is indeed invariant under $G^\sigma_{\rm GP}$. The limit $\tilde \phi_2 \to \infty$ amounts to imposing the base relation
\begin{equation} \label{eq:k=p2}
    k=p^2\,.
\end{equation}
The Weierstrass model over \begin{equation} \label{B30}
\cB_{3,0}\equiv\cB_{3,\tilde{\phi}_2\to\infty} \,
\end{equation}
 for an elliptic fibration over $\cB_{3,0}$ is then described by 
\begin{equation}
    y^2=x^3+f\,xz^4+g\,z^6\,,\qquad\Delta=4f^3+27g^2\,.
\end{equation}
Here, $f$ and $g$ are sections of $4\bar K_{\cB_{3,0}}$ and $6\bar K_{\cB_{3,0}}$ and therefore general polynomials of degrees 12 and 18 that are invariant under the action of $G_{\rm GP}$.
 
In Appendix~\ref{sapp:TypeOAEx} we derive the most general form of the sections $f$ and $g$, see in particular~\eqref{eq:P22233-34-generic-f} and~\eqref{eq:P22233-34-generic-g}. It then follows that for generic values of the coefficients of $f$ and $g$, the only singular fibers of the Weierstrass model are of Kodaira type I$_1$, over generic points of the discriminant divisor $\{\Delta =0\}$ (with vanishing orders ${\rm ord}(f,g, \Delta) = (0,0,1)$), or of Kodaira type II, over the curve  $f=g=0$ inside the discriminant divisor (with vanishing orders ${\rm ord}(f,g, \Delta) = (1,1,2)$). These are the generic singular fibers which every smooth Weierstrass model must have. While the fiber is singular, the total space of the fibration, and hence the elliptic fourfold, remains non-singular.
 
As there are no worse singular fibers, the F-theory fourfold $W$ is generically smooth even in the limit $\Tilde{\phi}_2\to\infty$. We conclude that this limit is indeed of type I$_{0,0}$ in the notation of \cite{Grimm:2019ixq}, which is a smooth locus at finite distance in the complex structure moduli space of $W$.\footnote{Clearly, additional tunings of the Weierstrass model can result in singularities. For example, setting $b_3 =0$ in (\ref{eq:P22233-34-generic-g}) gives rise to a smooth Weierstrass model away from $\hat \Delta_F$, while along $\hat \Delta_F$ the model exhibits a non-perturbative Kodaira type III fiber over the point $\{ f=0\} \cap \{ p=0\} \cap \{x_3=0\}$. In this case, the degeneration of the fourfold along $\hat \Delta_F$ is of type $I_{n,m}$ for $(n,m) \neq (0,0)$. Interestingly, the behaviour along $\hat \Delta_S$ is unaffected.}

This is indeed a rather drastic result: The F-theory uplift of the naive Type IIB infinite distance limit in which the Calabi--Yau threefold factorises corresponds to a specialisation in the F-theory moduli space which is not only at finite distance, but in fact completely non-singular. Responsible for this phenomenon are $g_s$ quantum effects away from weak coupling $g_s=0$ in F-theory.

 \paragraph{Behaviour along $\hat \Delta_S$.} 
  Having established that the Type IIB infinite distance degeneration is completely absent at generic values of $g_s$ in F-theory, we next turn to its realisation in the weak coupling regime, corresponding to taking
a Sen-limit. We first recall 
 how the Sen-limit works in the Weierstrass model starting from maximally generic functions $f$ and $g$ without enforcing $\tilde \phi_2 \to \infty$, i.e., at generic points along $\hat \Delta_S$.
A generic Sen-limit is obtained by parameterising the functions $f$ and $g$ of the Weierstrass model as \cite{Sen:1996vd}
\begin{equation} \label{eq:Senlimit-generic}
f = -3 h^2 + \epsilon \eta \,,  \qquad g = -2 h^3 + \epsilon h \eta - \frac{\epsilon^2}{12} \chi \,, \qquad 
\Longrightarrow \quad \Delta = 9 {\epsilon^2 } (h^2 (h\chi- \eta^2) + {\cal O}(\epsilon)) \,,
\end{equation}
with ${\rm deg}(h) = 6$, ${\rm deg}(\eta) = 12$, ${\rm deg}(\chi) = 18$, and then taking the limit $\epsilon \to 0$. For a generic choice of $h$, $\eta$, $\chi$, one recognizes the characteristic Kodaira I$_2$ degeneration of the general elliptic fiber in codimension-zero (because $(f,g, \Delta)$ have vanishing orders $(0,0,2)$ in $\epsilon$); this signals weak coupling globally over the base. Furthermore, the leading form of the discriminant shows that $h=0$ corresponds to the location of the O7-plane and $h\chi- \eta^2=0$ is the locus of the D7-branes.  In the model inherited from the specific orientifold under consideration, we recall from~\eqref{eq:Ofold2-Kbar} that the O7-plane is localised at $k=0$, and hence we set 
\begin{equation}
h=k \,,
\end{equation}
with $\eta, \chi$ generic, to describe the Sen-limit along a general point on 
$\hat\Delta_S$. 

Defining $(f_\epsilon,g_\epsilon,\Delta_\epsilon)$ to be the leading terms in the $\epsilon$-expansion of $(f,g,\Delta)$, we 
note the vanishing orders
\begin{equation} \label{232}
{\rm ord}(f_\epsilon,g_\epsilon,\Delta_\epsilon)\vert_{h=0}=(2,3,2)
\end{equation}
at the location of the O7-plane $h=0$ in the Sen-limit: These vanishing orders cannot occur in a generic Weierstrass model away from the Sen-limit, reflecting the famous effect that a naked O7-plane splits into two mutually non-local 7-branes in F-theory \cite{Sen:1996vd}. For $\epsilon =0$, such non-standard vanishing orders are possible because  the quantum effects (in $g_s$) responsible for this splitting of the O7-plane vanish here.

\paragraph{Intersection $\hat \Delta_F \cap \hat \Delta_{FS}$.}
With this preparation, let us now approach a Sen-type limit starting from
the smooth fourfold $W\to\cB_{3,0}$,
with $\cB_{3,0}$ as in \eqref{B30},
 i.e., on $\hat \Delta_F$. 
By first degenerating to $\tilde \phi_2 \to \infty$, and then enforcing a Sen-limit on top, we are in the growth sector where the $\tilde \phi_2 \to \infty$ limit is taken parametrically faster than the weak coupling limit. 
 To enforce a Sen-type limit over $\cB_{3,0}$, we must further specify 
 \begin{equation}
 h = k=p^2\,,
 \end{equation}
 which reflects the additional restriction that arises along the divisor $\hat \Delta_F$ in moduli space. This leads to the vanishing orders, in the sense explained before (\ref{232}),
\begin{equation} \label{orders-464}
    {\rm ord}(f_\epsilon,g_\epsilon,\Delta_\epsilon)\vert_{p=0}=(4,6,4)  \,.
\end{equation}
The doubling of vanishing orders compared to (\ref{232}) signals that the would-be orientifold locus becomes non-reduced as a consequence of $k=p^2$. In Type IIB language, this would lead to a stack of coincident O7-planes, which is a highly non-perturbative configuration. Indeed, from the perspective of the fivefold underlying the Sen-limit, the vanishing orders at $p=0$ are
\begin{equation}
    {\rm ord}(f,g,\Delta)\vert_{\epsilon=p=0}=(4,6,6)\,.
\end{equation}
The pathological nature of these vanishing orders becomes even clearer by noticing that the degeneration is related, via a simple base change $\epsilon \to \epsilon^4$, to a configuration with a non-minimal fiber in codimension-one:
Namely, after setting $h = p^2$ and replacing $\epsilon \to \epsilon^4$\footnote{While this is the minimal base change required to make the infinite nature of the singularity manifest, note that taking $\epsilon \to \epsilon^{4+n}$ for $n > 0$ gives a $(4,6, 12+2 n)$ singularity. The singularity is therefore strictly non-minimal (Class 4 in the nomenclature of \cite{Lee:2021qkx}).} in (\ref{eq:Senlimit-generic}) to implement this base change, 
\begin{eqnarray} \label{eq:Senlimit-F}
&& f = -3 p^4 + \epsilon^4 \eta \,,  \qquad g = -2 p^6 + \epsilon^4 p^2 \eta - \frac{\epsilon^8}{12} \chi \,, \\
&&\Delta = 9 \eta^2 p^4 \epsilon^8 + 9 \chi p^6 \epsilon^8 + 4 \eta^3 \epsilon^{12} - 
 \frac{9}{2} \chi \eta p^2 \epsilon^{12} + \frac{3}{16} \chi^2 \epsilon^{16} \,,
\end{eqnarray}
so that one finds the non-minimal vanishing orders 
\begin{equation}
    {\rm ord}(f,g,\Delta)\vert_{\epsilon=p=0}=(4,6,12)\,.
\end{equation}
Such non-minimal fibers in codimension-one and their physics interpretation have been studied in detail in \cite{Lee:2021qkx, Lee:2021usk} for elliptic K3 surfaces and in \cite{Alvarez-Garcia:2023gdd, Alvarez-Garcia:2023qqj} for elliptic threefolds. 
 They can be removed by 
 performing a sequence of blow-ups on the base.
 
In fact, the blow-up analysis performed in~\cite{Lee:2021qkx,Lee:2021usk} for elliptic K3 surfaces carries over, with important changes, to the example of $\cB_{3,0}$. A single blow-up is sufficient to resolve the non-minimal singularity in codimension one. 
 Parameterising the blow-up as 
\begin{equation}
\epsilon\mapsto e_0e_1 \,,  \qquad p\mapsto pe_1, 
\end{equation}
 and passing to the proper transform by dividing $(f,g,\Delta)$ by overall factors of $(e_1^4, e_1^6,e_1^{12})$ leads to the model
\begin{eqnarray}
&&f = - 3 p^4 +  e_0^4 \eta \,, \qquad g = - 2 p^6 + e_0^4  p^2 \eta  -\frac{1}{12}\chi e_0^8 e_1^2 \,,\label{SenlimitF1} \\
&&\Delta = \frac{1}{16} e_0^8 (3 \chi^2 e_0^8 e_1^4 + 64 e_0^4 \eta^3 - 72 \chi e_0^4 e_1^2 \eta p^2 - 
  144 \eta^2 p^4 + 144 \chi e_1^2 p^6) \label{SenlimitF2} \,.
\end{eqnarray}
At first sight, the base splits into two components parametrised by $\{e_0 =0\}$ and $\{e_1 =0\}$, with a generic $I_8$ fiber over  every point along the component $\{e_0 =0\}$ (because $({\rm ord}(f,g,\Delta)|_{e_0=0} = (0,0,8))$. 
 The degeneration of the generic fiber to I$_8$ indicates that the component $\{e_0 =0 \}$ is globally at weak coupling. 
 In the presence of general 7-branes, this would require infinite tuning and hence indicate an infinite distance degeneration.
 However, in the present model, this is not the case. The reason is that $(p,e_0)$ cannot vanish simultaneously after the blow-up, and therefore  $f$ and $g$ are constant and, in fact, non-vanishing along $\{e_0 =0\}$. Correspondingly, the fibration is trivial and only perturbative 7-branes (of A-type) are localised over the component $\{e_0=0\}$.\footnote{
 The location of the perturbative D7-branes on $e_0$ is given by $\Delta|_{e_0=0}$, i.e., by the intersection $\{e_0=0\}$ with  $\eta^2 - \chi e_1^2=0$. These branes are all mutually local with respect to one another and with respect to the duality frame in which $g_s \to 0$ on $\{e_0=0\}$. The perturbative value of $g_s$ at the location of these branes extends globally along the component $\{e_0 =0\}$ without extra tuning.} This shows that if one simply blows down this component by setting $e_0\equiv1$, these mutually local 7-branes  coincide, but without inducing any non-minimality or other pathology. Indeed, after this blow-down the base $\cB_{3,0}$ consists of the single component $\{e_1=0\}$ over which
\begin{equation}
    f=-3p^4+\eta\,,\quad g=-2p^6+p^2\eta\,,\quad \Delta=\eta^2(4\eta-9p^4)\,.
\end{equation}
In this model, the fiber over $\{p=0\}$ (where in the Type IIB orientifold the O7-plane is localised) is smooth, just as in the generic Weierstrass model away from the Sen-limit. The specialised Sen-limit (\ref{SenlimitF1}, \ref{SenlimitF2}) hence leads to a strongly coupled model after all, which is of degeneration type I at finite distance.\footnote{The fact that a non-minimal Kodaira singularity of $(4,6,12)$ can give rise to a finite distance type I model after passing to the semi-stable geometry may be surprising in view of the situation on K3 \cite{Lee:2021qkx, Lee:2021usk}, where such singularities are always at infinite distance. In the current example, the difference is a consequence of the fact that ${\rm deg}(f) = 4 {\rm deg}(p)$, which ultimately implies that along $\{e_0 = 0\}$ no mutually non-local 7-branes are localised and this component can hence be blown down. On K3, by contrast, an analogous tuning of the Weierstrass model would give rise to a Type III.a Kulikov model at infinite distance \cite{Lee:2021qkx, Lee:2021usk} with one strongly coupled region and one or several weakly coupled regions with generic I$_n$ fibers.\label{fn:K3-comparison}} Notice, however, that while at generic points along $\Hat{\Delta}_F$ the Weierstrass model is smooth, at the intersection $\Hat{\Delta}_F\cap\Hat{\Delta}_{FS}$ there are codimension one I$_2$-fibers over $\{\eta=0\}$.\footnote{These are the net effect of blowing down the component $\{e_0=0\}$, which after all does carry mutually local perturbative D7-branes.} Therefore, this intersection of divisors is still at finite distance in moduli space, but the fourfold is not smooth anymore. In other words, the secondary singularity type has increased from $(0,0)$ to $(a,b)\neq(0,0)$. This shows furthermore that the resolution divisor $\Hat{\Delta}_{FS}$ has primary singularity type I.

\paragraph{Intersection $\hat \Delta_S \cap \hat \Delta_{FS}$.}
To determine the geometry at the intersection $\hat{\Delta}_S \cap \hat \Delta_{FS}$, we now reverse the orders of the Sen-limit versus the base complex structure degeneration, i.e., we first take the Sen-limit (\ref{eq:Senlimit-generic}) to arrive at a generic point on $\hat \Delta_S$ and then specialise further. The defining equation for the base $\cB_3$ can be written as (cf. \eqref{basedef}),
\begin{equation} \label{obenlinks}
k = p^2 +  u P_6  \,,  \qquad u = 1/{\sqrt{\tilde \phi_2}} \,, \qquad  P_6 = \sqrt{Q_{12}}.
\end{equation}
 On $\hat \Delta_S$, we set $\epsilon =0$, and therefore approaching $\hat \Delta_S \cap \hat \Delta_{FS}$ along $\hat \Delta_S$ amounts to the limit $u \to 0$ in
 \begin{equation}
f = -3 (p^2 +  u P_6)^2 \,,  \qquad g = -2 (p^2 +  u P_6)^3  \,, \qquad  \Delta \equiv 0. 
 \end{equation}
This model is not yet semi-stable, as can be seen from the fact that a base change $u \to u^2$ leads to non-minimal vanishing orders 
\begin{equation}
    {\rm ord}(f,g,\Delta)\vert_{u=p=0}=(4,6,12)\,.
\end{equation}
To remedy this, we again blow up the base via $u \to e_0 e_1$, $p \to p e_1$, and note that as before the component $\{e_0=0\}$ can be blown down. After this blow-down, the resulting Weierstrass model reads
\begin{equation} \label{model-semistable-final}
 f = -3 (p^2 + P_6)^2 \,, \qquad  g = -2 (p^2 + P_6)^3 \,, \qquad \Delta \equiv 0 \,.
\end{equation}
The overall vanishing of the discriminant is a consequence of setting $\epsilon \equiv 0$ and reflects the weak coupling nature of the limit. The generic fiber is of Kodaira type I$_n$ and consequently the fourfold degeneration  remains of type II even at the point $\hat \Delta_S \cap \hat \Delta_{FS}$. This can also be inferred from the observation that the Calabi--Yau threefold obtained in the Sen-limit as the double cover $\{k\equiv h=\xi^2\}$ does not factorise at the intersection $\Hat{\Delta}_S\cap\hat{\Delta}_{FS}$ once the non-minimal fibers are resolved. 

This may again seem counter-intuitive at first: From the point of view of the original Type IIB orientifold, one might have expected that the degeneration enhances from type II to type III because of the splitting of the original Calabi--Yau threefold in the limit $\phi_2\to\infty$. More precisely, in the generic Sen-limit the two components of the fourfold intersect along the Calabi--Yau double cover $\{h=\xi^2\}$ of the base $\cB_3$, which coincides with the original Calabi--Yau threefold that factorises in the infinite distance limit $\phi_2\to\infty$ in Type IIB. However, for an orientifold of O-type A, like the one under consideration here, the O7-plane is localised along the resulting double surface. From the perspective of the fourfold, this would give rise to a degeneration that is not semi-stable. This is taken care of automatically by the algorithm that passes to the semi-stable form \eqref{model-semistable-final}, which however is of type II and not of type III. In other words, the split of the Type IIB Calabi--Yau threefold in the limit $\phi_2 \to \infty$ has no analogue in F-theory, not even at $g_s =0$, i.e., at the point $\hat \Delta_S \cap \hat \Delta_{FS}$. Notice that this difference between the singularity type expected from Type IIB perspective and the actual singularity type in F-theory provides a very violent example for the physical interpretation of F-theory lifts of O-type A orientifolds given in Section~\ref{ssec:general}.

\paragraph{The divisor $\hat \Delta_{FS}$.}
Finally, along the locus $\hat \Delta_{FS}$ in complex structure moduli space, one can interpolate between the Weierstrass models  
(\ref{eq:Senlimit-F}) and (\ref{obenlinks}). As seen in the discussion of the intersection $\Hat{\Delta}_F\cap\Hat{\Delta}_{FS}$ above, the degeneration is of type I$_{a,b}$ and only enhances to type II at $\hat \Delta_S \cap \hat \Delta_{FS}$.

\subsection{O-Type A orientifolds with a single O7-plane} 

In the previous section, we considered a specific example of an O-type A infinite distance limit in a Type IIB orientifold and its uplift to F-theory.
  The conclusion 
 of this discussion holds more generally  
  for orientifolds of O-type A containing only a single O7-plane: Generically, 
  the number of components into which the 
  double cover of the base factorises at    
$\Hat{\Delta}_S\cap\Hat{\Delta}_{FS}$
  is reduced compared to the naive Type IIB orientifold expectation, reflecting a strong deviation from the perturbative picture.
  
 \paragraph{Behaviour at $\hat \Delta_F$.}  The F-theory uplift of a generic O-Type A model with a single O7-plane divisor can be constructed as in Section~\ref{ssec:Ftheorylift}.
 The O7-plane locus lifts, in F-theory, to a single irreducible divisor, $\{k=0\}$, and $k$ enters the defining equation of $\cB_3$ in the combination 
  \eqref{degenerationOA}.
 Along generic points of $\hat \Delta_F$, the base therefore splits into the components
 \begin{equation}
{\cal B}_{3,0} = V_{p^2-k} \, \cup  \, \bigcup_{j'} {V}_{j'} \,.
 \end{equation}
The base components 
${V}_{j'}$ are not affected by the factorisation of the O7-plane along $V_{p^2-k}$ and thus are simply spectators. 
The total number of base components in
O-type A limits is smaller by at least one compared to the naive Type IIB limit. 
 As we have seen in the examples, this reduction may even decrease the primary type of degeneration along $\hat \Delta_F$ compared to the Type IIB expectation, but it does not have to.

\paragraph{Behaviour at $\hat \Delta_S \cap \hat{\Delta}_{FS}$.} Taking first the Sen-limit and then further specifying the Weierstrass sections $f$ and $g$ similar to~\eqref{obenlinks}
  ultimately leads to non-minimal fibers in codimension-one over the base component
  $V_{p^2-k}$, while the fibers over the components ${V}_{j'}$ become generically of type I$_{n>0}$.
 Resolving the non-minimality by blow-up and blow-down
 leads to the same total number of base components at $\Hat{\Delta}_S\cap\Hat{\Delta}_{FS}$
  as over generic points of 
$\hat \Delta_F$. Indeed, the blow-down, as in the example of the previous section, is always possible as long as there is only one O7-plane divisor because in this case $f$ and $g$ along the blow-up component $\{e_0 = 0\}$ of the base are trivial.\footnote{In view of Footnote~\ref{fn:K3-comparison}, we expect that for F-theory lifts of O-type A orientifolds with more than one O7-plane there can indeed be an enhancement of the number of components at the intersection $\Hat{\Delta}_S\cap\Hat{\Delta}_{FS}$ compared to $\Hat{\Delta}_F$.}

Altogether, one ends up with the following 
 fourfold at $\Hat{\Delta}_S\cap\Hat{\Delta}_{FS}$:
 The double cover of the base, identified with the Type IIB Calabi--Yau threefold,
 degenerates into the same number of components as the base, and in particular 
 into fewer components than expected from the naive Type IIB orientifold perspective.
   At the intersection of 
  $\hat \Delta_{FS} \cap \hat \Delta_S$, the degeneration is then of type 
   \begin{equation}
\mathtt{A}_{\hat \Delta_S \cap \hat \Delta_{FS}} = \mathtt{A}_{\hat \Delta_F} + \mathtt{I}\,.
   \end{equation}
Recall that the naive expectation would have been that always 
$\mathtt{A}_{\hat \Delta_S \cap \hat \Delta_{FS}} = \mathtt{A}_{\rm IIB} + \mathtt{I}$.
As we have seen, at least for orientifolds with a single O7-plane, this is not necessarily the case. 
 Indeed, in Section~\ref{ssec:Ftheorylift} we have found examples with the following configurations of singularity types: 
 \begin{center}
\begin{tabular}{|c|c|c|}
 \hline
     $\mathtt{A}_{\rm IIB}$& $\mathtt{A}_{\hat{\Delta}_F}$ & $\mathtt{A}_{\hat{\Delta}_S\cap \hat{\Delta}_{FS}}$ \\  [0.5ex] \hline  
     II &  I & II \\ 
     III & II & III \\ 
     IV & IV &V \\\hline
\end{tabular}
\end{center}

\paragraph{Behaviour at $\hat \Delta_F \cap \hat{\Delta}_{FS}$.} 
Along the base component $V_{p^2-k}$,
 the effect of taking the Sen-limit after first factorising the base 
is completely analogous to the discussion in the previous section and leads to a component at strong coupling; by contrast, the generic fibers over the additional base components $V_{j'}$, if there are any, degenerate to Kodaira I$_n$ type as a result of the Sen-limit. 
 The model is then locally weakly coupled over $V_{j'}$ but not globally due to the strong coupling behaviour along $V_{p^2 -k}$.
 This enhances the primary type of the degeneration to
\begin{equation}
\mathtt{A}_{\hat \Delta_F \cap \hat \Delta_{FS}} = 
\begin{cases}
        \mathtt{A}_{\hat \Delta_F} + \mathtt{I} & {\rm if}\quad \mathtt{A}_{\hat \Delta_F} \geq \mathtt{II} \, \\
         \mathtt{A}_{\hat \Delta_F}   & {\rm if}\quad \mathtt{A}_{\hat \Delta_F} = \mathtt{I} \,. \\
      \end{cases}
   \end{equation}
Furthermore, the singularity type along 
 $\hat \Delta_{FS}$ remains the same until one reaches the intersection with $\hat \Delta_S$.

\subsection{Comparison to the moduli space of Calabi--Yau threefolds}

The properties of the Type IIB quantum moduli space described above may seem quite radical from the perturbative Type IIB perspective: After all, we claim that O-type A limits cannot be described within the regime of validity of the perturbative Type IIB effective action. However, from the perspective of the Calabi--Yau fourfold we are still dealing with the classical complex structure moduli space. The fact that the actual type of a singularity can differ from the classical expectation already arises in moduli spaces of Calabi--Yau threefolds. It is instructive to consider such a Calabi--Yau threefold example and provide the dictionary to relate our proposal above to the better understood moduli spaces of Calabi--Yau threefolds. 

\begin{figure}
    \centering
    \includegraphics[width=0.5\linewidth]{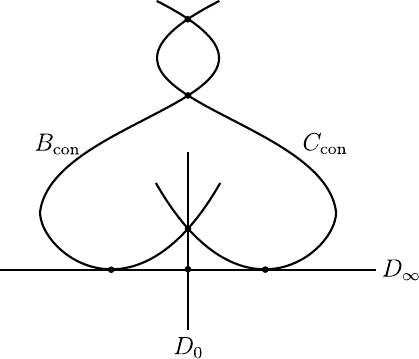}
    \caption{A sketch of the discriminant locus in the complex structure moduli space of the mirror of $\mathbb{P}^4_{1,1,1,6,9}[18]$. Reproduced from Figure 3.1 in \cite{Candelas:1994hw}.}
    \label{fig:11169}
\end{figure}

As an example, consider Type IIA string theory compactified on $\hat{V}=\mathbb{P}^4_{1,1,1,6,9}[18]$.\footnote{Equivalently, we could consider Type IIB string theory on the mirror manifold. However, in the following we use the properties of the vector multiplet moduli space in the large volume regime of $\hat{V}$ which is easier to describe in Type IIA language.} We focus on the K\"ahler moduli spanning the vector multiplet sector of the 4d $\cN=2$ effective theory. This moduli space is complex two-dimensional and was first discussed in detail in~\cite{Candelas:1994hw}. After resolving the orbifold singularities in the ambient space $\mathbb{P}^4_{1,1,1,6,9}$, the manifold $\hat{V}$ can be seen as an elliptic fibration over $\mathbb{P}^2$. The two K\"ahler moduli are the volume $t_b$ of the hyperplane class $H$ in the $\mathbb{P}^2$ base and the volume $t_f$ of the elliptic fiber. As analysed in~\cite{Candelas:1994hw}, the naive moduli space of the theory has only four special divisors (see Figure \ref{fig:11169}): an orbifold divisor $D_0$, two conifold divisors $B_{\rm con}$ and $C_{\rm con}$ and an infinite distance divisor $D_\infty$ of type III. For the analogy to the F-theory discussion above, only $D_\infty$ and $B_{\rm con}$ are relevant. As is clear from the discussion in~\cite{Candelas:1994hw}, these two divisors intersect in a point of tangency. Resolving this point of tangency introduces a number of additional divisors. Of relevance for us is the type IV divisor denoted by $G_1$ in~\cite{Candelas:1994hw} which intersects $D_\infty$ and $B_{\rm con}$ normally. Comparing with the set of divisors in the Type IIB/F-theory discussion of Section~\ref{ssec:general}, we can make the following identification:
\begin{center}
    \begin{tabular}{c c c}
         F-theory/Type IIB & $\longleftrightarrow $ & Type IIA on $\mathbb{P}_{1,1,1,6,9}[18]$ \\\hline \hline
         Sen-Limit Divisor $\hat{\Delta}_S$&\multirow{3}{*}{$\longleftrightarrow$}&  Type III Divisor $D_{\infty}$ \\  
         O-Type A Divisor $\hat{\Delta}_F$ &  & Conifold Divisor $B_{\rm con}$\\  
         Blow-up Divisor $\hat\Delta_{FS}$&& Blow-up Divisor $G_1$\\
    \end{tabular}  
\end{center}
Notice that here we have a type III divisor $D_\infty$ whereas the Sen-limit divisor $\hat{\Delta}_S$ is of type II. We now argue that the conifold divisor $B_{\rm con}$ can be viewed as the quantum version of what ``classically'' one would have expected to be a type II divisor. For this reason, $B_{\rm con}$ can indeed be viewed as an O-type A divisor with $\mathtt{A}={\rm II}$. 

To appreciate the analogy to the Type IIA string on $\mathbb{P}^4_{1,1,1,6,9}[18]$, we first have to identify the ``classical'' perturbative theory that describes the theory in the vicinity of the divisor $D_{\infty}$. This will be the analogue of the perturbative Type IIB orientifold theory in the F-theory description above. In terms of the K\"ahler moduli $(t_b, t_f)$, the divisor $D_\infty$ corresponds to $t_b\to \infty$. This limit corresponds to a decompactification limit to 6d. The classical, perturbative theory thus corresponds to F-theory on $\hat{V}\times T^2$ where the torus factor is made large. Dimensionally reducing the 6d effective action on the torus leads to a classical K\"ahler potential 
\begin{equation}\label{Kclass6dto4d}
    e^{-K_{6d\to 4d}} = \frac{1}{2} t_f t_b^2\,. 
\end{equation}
At this level, the moduli space seems to factorise between the fibral and base K\"ahler moduli. This is the analogue of the perturbative factorisation of the Type IIB moduli space in the Sen-limit. Within the perturbative description, we could now consider the limit $t_f\to \infty$. Based on the classical K\"ahler potential in~\eqref{Kclass6dto4d}, we would conclude that this corresponds to a type II divisor. However, we know that there is no type II divisor in the moduli space of $\hat{V}$. Instead the $t_f\to \infty$ limit can only be identified with the conifold divisor $B_{\rm con}$ which is of type I. This is analogous to the reduction of the singularity type of $\Delta_F$ in F-theory. Along $B_{\rm con}$ the central charge of the D4-brane wrapping the zero section $S_0$ of the elliptic fibration of $\hat{V}$ vanishes. The D4-brane on $S_0$ is non-perturbative from the perspective of the 6d theory.\footnote{This can be seen in the dual M-theory compactification on $\hat{V}\times S^1$. Here, the D4-brane maps to an M5-brane wrapping $S_0\times S^1$. The 5d string obtained by wrapping the M5-brane on $S_0$ becomes tensionless at finite distance in the 5d vector multiplet moduli space signalling the appearance of a 5d SCFT. The 5d SCFT locus in the M-theory vector multiplet moduli space is realized at finite volume for the elliptic fiber of $\hat{V}$, $t_f\sim \mathcal{O}(1)$. Using the M-/F-theory dictionary, $t_f\sim \mathcal{O}(1)$ corresponds to a string-size radius for the extra F-theory circle. Thus, the M5-brane on $S_0$ is non-perturbative from the perspective of the 6d theory and so is the D4-brane in Type IIA.} This fact is reflected in the point of tangency between $D_\infty$ and $B_{\rm con}$ prior to resolution and is therefore analogous to the point of tangency between $\Delta_F$ and $\Delta_S$ due to the non-perturbative nature of coincident O7-planes from the Type IIB perspective.

Notice that in the vicinity of the intersection $G_1\cap D_\infty$ the theory does not simply correspond to a 6d theory compactified on a $T^2$ but is a genuine 4d theory. For this reason, the effective action derived from the compactification of the 6d theory to 4d as determined by the K\"ahler potential~\eqref{Kclass6dto4d} is not valid in this regime. This is exactly analogous to our discussion of the fourfold moduli space. In that case, it is the perturbative Type IIB orientifold effective action that is not valid in the vicinity of the locus $\hat{\Delta}_{FS}$. 

The comparison with the well-understood K\"ahler moduli space of $\mathbb{P}^4_{1,1,1,6,9}[18]$ demonstrates that our results for the quantum moduli space in the vicinity of O-type A limits are quite standard from the perspective of Calabi--Yau moduli spaces. It further shows that trusting the perturbative Type IIB orientifold effective action away from generic points on $\Delta_S$ is equivalent to describing the K\"ahler moduli space of an elliptically fibered Calabi--Yau threefold by completely forgetting the non-trivial twist of the fiber, i.e., ignoring the fact that supersymmetry is broken from $\cN=4$ to $\cN=2$. The comparison with Type IIA illustrates yet another point: Even though our discussion focused mainly on infinite distance regimes, these are just the regimes where the mismatch between the perturbative effective action and the actual theory becomes clearest. However, even in the interior of the moduli space of $\hat{V}$ one would not trust the K\"ahler potential~\eqref{Kclass6dto4d} to describe the effective action. It is only in the parametric regime $t_b\gg t_f\gg 1$ that the K\"ahler potential in~\eqref{Kclass6dto4d} is a good approximation to the real K\"ahler potential. Applied to the analogous Type IIB orientifolds, one should trust the 4d $\cN=1$ effective action derived from perturbative Type IIB string theory just in the parametric regime $|\tau|\gg |z_B|$, $|z_B|<\infty$ for all complex structure moduli $z_B$ that classically yield O-type A limits if sent to infinity.

\section{Conclusions}\label{sec:conclusions}
This article is the first part of our analysis of infinite distance limits in field spaces of four-dimensional $\cN=1$ compactifications of Type IIB/F-theory. We have investigated whether the perturbative Type IIB effective action correctly describes the physics of infinite distance limits in the classical complex structure moduli space of orientifolds of Calabi--Yau threefolds. Our results demonstrate a rather striking mismatch between the perturbative Type IIB moduli space and the actual $g_s$-corrected moduli space of the full theory of quantum gravity. The main take-aways of our analysis can be summarised as follows:
\begin{itemize}
    \item The properties of the quantum gravitational theory are encoded in the details of the \emph{geometry} of the Calabi--Yau threefold $V$ whose orientifold is taken. From the low-energy supergravity approximation it is sufficient to check whether a complex structure deformation is associated with a 3-form in $H^3_-(V/\Omega)$ to determine whether it survives as a scalar mode in the 4d $\cN=1$ theory. By contrast, in a given asymptotic limit of $\cM_{\rm c.s.}(V/\Omega)$ the position of the O7-plane relative to the asymptotic degeneration of $V/\Omega$ controls whether the limit receives strong $g_s$ corrections. This exemplifies yet again that in string theory compactifications, genuine quantum gravitational properties of the theory are encoded in the details of the geometry and may not be directly accessible from the dimensionally reduced effective action. 
    \item Geometrically, the infinite distance limits in the complex structure limit are characterised by a multi-component normal crossing variety $V_0$ into which the Calabi--Yau threefold splits. If after orientifolding, the O7-plane has support on the intersection of two components of $V_0$, we call the corresponding limit in $\cM_{\rm c.s.}(V/\Omega)$ to be of \emph{O-type A}. These limits receive strong $g_s$ corrections which can be attributed to the fact that the O7-plane divisor factorises in these limits. 
    \item The F-theory uplift reveals that the singularity type of an O-type A limit can change due to $g_s$ corrections. Moreover, the original limit in the closed Type IIB moduli space becomes a mixed open-closed string limit in the $g_s$-corrected moduli space. This, again, is clear from the splitting of the O7-plane locus. 
    \item The divisor in the F-theory moduli space corresponding to the uplift of the O-type A limit has a point of tangency with the divisor associated with the Sen-limit, i.e., the weak-coupling regime for the Type IIB string. This point of tangency signals that non-perturbative effects are important even in the weak-coupling limit. Physically, this means that even in the weak-coupling limit perturbative control over the Type IIB effective action is lost if also an O-type A limit in $\cM_{\rm c.s.}(V/\Omega)$ is imposed. The reason is, again, that in the O-type A limit two O7-planes coincide, which is a highly non-perturbative configuration from a Type IIB orientifold perspective. This behaviour of the $g_s$-corrected moduli space illustrates that the existence of a weak string-coupling limit need not imply that there is a well-defined perturbative string expansion. This is well-known from heterotic compactifications in the presence of NS5-branes and has recently been stressed again in~\cite{Monnee:2025msf}. 
\end{itemize}
Our results show that the $g_s$-corrected effective action of 4d $\cN=1$ theories obtained from Type IIB orientifolds differs significantly from the classical effective theory. In particular, a large class of infinite distance regimes in the classical moduli space $\cM_{\rm c.s.}(V/\Omega)$ lie outside the regime of validity of the perturbative Type IIB effective action. In contrast, the classical F-theory complex structure moduli space encodes all pure $g_s$ corrections. It is in fact a common feature of quantum gravity that quantum effects of a given perturbative description can be encoded in classical quantities in a dual frame. 

A natural next question is whether the classical F-theory effective action receives further corrections which, from the perturbative Type IIB point of view, correspond to $\a'$ and mixed $g_s, \alpha'$ corrections. The perhaps surprising effect of these corrections on asymptotic regimes in the F-theory complex structure moduli space and its consequences for model building is the subject of the companion paper~\cite{Paper2}.

\subsubsection*{Acknowledgements}
We thank Bj\"orn Hassfeld, Arthur Hebecker, Seung-Joo Lee, Severin L\"ust and Luca Martucci for useful discussions. JM thanks the string theory group at IFT, Madrid, for hospitality and interesting discussions. This work is supported in part by Deutsche Forschungsgemeinschaft under Germany’s Excellence Strategy EXC 2121 Quantum Universe 390833306, by Deutsche Forschungsgemeinschaft through a Ger\-man-Israeli Project Cooperation (DIP) grant “Holography and the Swampland” and by Deut\-sche Forschungsgemeinschaft through the Collaborative Research Center 1624 “Higher Structures, Moduli Spaces and Integrability.” MW acknowledges support by  Deutsche Forschungsgemeinschaft through the Emmy Noether program 557478919.

\appendix
\section{Instantons in Type I compactifications}\label{app:TypeI-D1}
In this appendix we give more details on the perturbative and non-perturbative corrections to the classical K\"ahler potential of Type I string theory compactified on $V={\rm K3}\times T^2$ which is relevant for the discussion in Section~\ref{ssec:simple-ex}. Recall from~\eqref{eq:K-corr-TypeI} that the corrected K\"ahler potential takes the form
\begin{equation}
    K=-\log\left[(S-\bar{S})(T-\bar{T})(U-\bar{U})\right]+\frac{\ii}{2}\frac{V_{\rm 1-loop}+V_{\rm D1}}{S-\bar{S}}.
\end{equation}
The one-loop contribution has been computed in~\cite{Antoniadis:1996vw,Berg:2005ja} as
\begin{equation}
    V_{\rm 1-loop}=-\frac{4\pi\ii}{3}\frac{E(U_I,2)}{T-\bar{T}},
\end{equation}
with the non-holomorphic Eisenstein series
\begin{equation}
    E(U,k)=\frac{1}{\zeta(2k)}\sum_{(j_1,j_2)\neq(0,0)}\frac{({\rm Im}(U))^k}{|j_1+j_2U|^{2k}}.
\end{equation}
For us, the non-perturbative corrections due to D1-instantons are more relevant. These have been computed in~\cite{Camara:2008zk} using the dual heterotic description of the theory.  The D1-instantons in Type I theory become worldsheet instantons which are encoded in the perturbative heterotic partition function. As toroidal compactifications of the heterotic string enjoy modular invariance, the D1-instanton correction can be expressed in terms of modular functions as 
\begin{equation}
    V_{\rm D1}=-\frac{1}{\pi}\sum_{k>j\geq0,\,p>0}\frac{e^{2\pi\ii kpT}}{(kp)^2}\left(\frac{\Hat{A}_K(\cU)}{{\rm Im}(T)}-\frac{2\pi\ii kp}{\cU-\bar{\cU}}\frac{E_{10}(\cU)}{\eta^{24}(\cU)}\right)+{\rm c.c.}\,.
\end{equation}
The complex parameter $\mathcal{U}$, defined as
\begin{equation}\label{eq:D1-cU2}
    \cU=\frac{j+pU_I}{k}\,,
\end{equation}
encodes the wrapping of the D1-brane along the torus which is specified by the three integers $j,p,k$ over which the sum in $V_{\rm D1}$ is taken. The function $\Hat{A}_K(\cU)$ is defined as
\begin{equation}
    \Hat{A}_K(\cU)=\frac{\Hat{E}_2E_4E_6+2E_6^2+3E_4^3}{12\eta^{24}}
\end{equation}
with all Eisenstein series and the Dedekind eta function evaluated at $\cU$ as given in \eqref{eq:D1-cU2}. Also, $\Hat{E}_2=E_2-\frac{3}{\pi{\rm Im}(\cU)}$. From this it is evident that the one-loop determinant in the parantheses in $V_{\rm D1}$ is modular invariant with respect to $\cU$. In fact, as was noted already in~\cite{Camara:2008zk}, $V_{\rm D1}$ can be written in terms of degree-zero Hecke operators as
\begin{equation}\label{eq:D1-modular}
    V_{\rm D1}=-\frac{1}{2\pi}\sum_{n=1}^\infty e^{2\pi i n T}H_n[\Phi_n](U)+{\rm c.c.}\,,
\end{equation}
where
\begin{equation}
    H_n[\Phi_n](U)=\frac{1}{n}\sum_{p>0,\,kp=n}\sum_{k>j\geq0}\Phi_n\left[\frac{j+pU}{k}\right],\quad \Phi_n(\cU)=\frac{\Hat{A}_K(\cU)}{n{\rm Im}(T)}-\frac{2\ii}{\cU-\bar{\cU}}\frac{E_{10}(\cU)}{\eta^{24}(\cU)}\,,
\end{equation}
showing that $V_{\rm D1}$ is modular invariant also in the $T^2$-complex structure modulus $U_I$.

We are interested in the behaviour of $V_{\rm D1}$ in the limit $U_I\to\ii\infty$, i.e., $q=e^{2\pi\ii\cU}\to0$. The one-loop determinant can then be expanded as
\begin{equation}
    \left(\frac{\Hat{A}_K(\mathcal{U})}{{\rm Im}(T)}-\frac{2\pi\ii kp}{\mathcal{U}-\bar{\mathcal{U}}}\frac{E_{10}(\mathcal{U})}{\eta^{24}(\mathcal{U})}\right)=\sum_{m=-1}^\infty\left(\frac{a_m}{{\rm Im}(T)}-\frac{k^2}{{\rm Im}(U_I)}b_m\right)q^m\,.
\end{equation}
The only $j$-dependence in this expression is through $q$. Notice that the sum over $j=0,..,k-1$ of $q^m$ for $m\neq0$ vanishes unless $k=1$. The leading term in the $q\to0$ limit comes from $m=-1$, which is why we focus on this term in the following. To determine the relevant coefficients $a_{-1}$ and $b_{-1}$, recall the following small $q$-expansions,
\begin{align}
    \begin{split}
        E_2(q)&=1-24q+\mathcal{O}(q^2)\,,\\
        E_4(q)&=1+240q++\mathcal{O}(q^2)\,,\\
        E_6(q)&=1-504q+\mathcal{O}(q^2)\,,\\
        E_{10}(q)=E_4(q)E_6(q)&=1-264q+\mathcal{O}(q^2)\,,\\
        \eta^{24}(q)&=q(1-24q+\mathcal{O}(q^2))\,.
    \end{split}
\end{align}
Using this, we find
\begin{equation}
    \hat{A}_K(\mathcal{U})=\left(\frac{1}{2}-\frac{1}{4\pi\,\mathrm{Im}(\mathcal{U})}\right)q^{-1}+\mathcal{O}(q^0),\quad\frac{E_{10}(\mathcal{U})}{\eta^{24}(\mathcal{U})}=q^{-1}+\mathcal{O}(q^0)\,,
\end{equation}
i.e.
\begin{equation}
    a_{-1}=\frac{1}{2}-\frac{1}{4\pi p\,\mathrm{Im}(U)}\,,\quad
    b_{-1}=1\,,
\end{equation}
where for $a_{-1}$ we used that $k=1$. With $q=\exp(2\pi i p U_I)$ we can write
\begin{equation}
    V_{\rm D1}=-\frac{1}{\pi}\sum_{p=1}^\infty\frac{e^{2\pi \ii p\,T}}{p^2}\left(\frac{1}{\mathrm{Im}(T)}\left(\frac{1}{2}-\frac{1}{4\pi p\,\mathrm{Im}(U)}\right)-\frac{1}{\mathrm{Im}(U)}\right)e^{-2\pi i p U}+\cO(q^0) + {\rm c.c.}\,,
\end{equation}
which is the form given in~\eqref{eq:VD1-largeU}. 

As discussed in more detail in~\ref{ssec:simple-ex}, to maintain perturbative control in $g_s$ in the Type IIB orientifold, one needs to superimpose the limit $U_A\to\ii\infty$ with the limit $g_s^{-1}\sim\Im(U_A)\to\infty$. However, even then there are further corrections due to D3-brane instantons, which correspond to D$5$-brane instantons wrapping the full internal space in Type I. Note that these corrections have not been computed in~\cite{Camara:2008zk}. Still, we can make use of the extended supersymmetry of this example and use duality of the Type I setup to Type IIA string theory compactified on the smooth Calabi--Yau threefold $V_{\rm IIA}=\mathbb{P}^4_{1,1,2,8,12}[24]$~\cite{Antoniadis:1996vw}. The threefold $V_{\rm IIA}$ can be viewed as an elliptic fibration over $\mathbb{F}_0=\mathbb{P}^1_f\times \mathbb{P}^1_b$ which thus has $h^{1,1}=3$. Type IIA on this threefold gives rise to a 4d $\cN=2$ theory with three vector multiplets with scalar components given by the complexified K\"ahler moduli of $V_{\rm IIA}$ which we denote by 
\begin{equation}
    t_U = \int\limits_{\cE}\left(B_2+iJ\right) \,,\quad t_f = \int\limits_{\mathbb{P}^1_f}\left(B_2+iJ\right) \,,\quad  t_b = \int\limits_{\mathbb{P}^1_b}\left(B_2+iJ\right)\,, 
\end{equation}
where $J$ is the K\"ahler form on $V_{\rm IIA}$ and $\cE$ the elliptic fiber of $V_{\rm IIA}$. The classical prepotential for the vector multiplet sector of the 4d $\cN=2$ theory is given by 
\begin{equation}
    F_{\rm IIA} = t_U t_f t_b + t_U^2(t_f+t_b) + \frac{4}{3} t_U^3\,. 
\end{equation}
To match the moduli $t_{U,b,f}$ with the moduli $(S,T,U)$ in the Type I duality frame, we notice that the perturbative K\"ahler potential in~\eqref{eq:K-corr-TypeI} is an STU-model with a one-loop correction that only depends on the modulus $U$. This structure can be reproduced in the Type IIA formulation of the theory by identifying (see also~\cite{Berglund:2005dm})
\begin{equation}
    t_U = U \,,\qquad t_f = T - U \,,\qquad t_b=S-U\,, 
\end{equation}
which yields the prepotential 
\begin{equation}
    F_{\rm IIA} = S T U +\frac13 U^3\,,
\end{equation}
from which an STU-type K\"ahler potential can be derived with a correction that is only $U$-dependent. Comparing with~\eqref{eq:SD(-1)-corr}, we observe that $t_f$ expressed in terms of $T$ and $U$ agrees with the corrected D1-brane instanton action in the Type I duality frame. In the Type IIA frame, $t_f$ is the action of a worldsheet instanton wrapping $\mathbb{P}^1_f$. From this perspective it is not surprising that in the limit $U\to \ii \infty$, at constant $T$, instanton actions become unsuppressed. This is because in the Calabi--Yau compactification of Type IIA string theory, this limit corresponds to shrinking the curve $\mathbb{P}^1_f$. 

From this dual formulation, we further infer that there is an exchange symmetry $S\leftrightarrow T$. For $V_{\rm II A}$ this symmetry is manifest geometrically since the two $\mathbb{P}^1$s of $\mathbb{F}_0=\mathbb{P}^1_f \times \mathbb{P}^1_b$ are equivalent and the symmetry $S \leftrightarrow T$ holds also at the quantum level. In particular, the worldsheet instantons wrapping $\mathbb{P}^1_b$ are equivalent to those wrapping $\mathbb{P}^1_f$ under this symmetry. The worldsheet instanton wrapping $\mathbb{P}^1_b$ map under duality to Type I string theory to D5-brane instantons wrapping K3$\times T^2$. The exchange symmetry $\mathbb{P}^1_b\leftrightarrow \mathbb{P}^1_f$ enables us to infer the correction to the K\"ahler potential due to D5-brane instantons in the Type I duality frame which takes the form 
\begin{equation}\label{eq:kahlernonpert}
    K=-\log\left[\left(\mathrm{Im}(S)+V_{\rm D1}\right)(\mathrm{Im}(T)+V_{\rm D5})\mathrm{Im}(U)
    \left(1-\frac{V_{\rm 1-loop}}{(\mathrm{Im}(S)+V_{\rm D1})(\mathrm{Im}(T)+V_{\rm D5})}\right)\right]\,,
\end{equation}
where $V_{\rm D5}$ is the contribution from pure D5-brane instanton, i.e., instantons without induced D1-brane charge. Exploiting $S\leftrightarrow T$ symmetry of the theory, we find 
\begin{equation}
    V_{\rm D5} = -\frac{1}{\pi}\sum_{p=1}^\infty\frac{e^{2\pi \ii p\,S}}{p^2}\left(
    \frac{1}{\mathrm{Im}(S)}\left(\frac{1}{2}-\frac{1}{4\pi p\,\mathrm{Im}(U_I)}\right)
    -\frac{1}{\mathrm{Im}(U_I)}\right)e^{-2\pi i p U_I}+\cO(q^0) + {\rm c.c.}\,,
    \label{eq:TypeI-D5-series}
\end{equation}
where we used a similar expansion as in~\eqref{eq:VD1-largeU}. In particular, this means that the D5-brane instanton action also becomes unsuppressed in the limit $U_I\to\ii \infty$ at constant $S$. Performing two T-dualities along $T^2$ to the O-type A setup, the D5-brane instantons become D3-brane instantons wrapping the K3-factor which, hence, become unsuppressed in the $U\to \ii \infty$ limit.\footnote{Similarly, assuming the K3 to be elliptically fibered, performing two T-dualities along the generic fiber maps these D5-brane instantons to D3-brane instantons wrapping the elliptic fiber and the $T^2$ in the O-type B setup discussed in \cite{Paper2}.} Thus, apart from the D$(-1)$-instantons that are geometrised in the F-theory lift, there are D3-brane instantons that become unsuppressed in the naive large complex structure limit. In addition, there are infinitely many instantons carrying both D3- and D(-1)-brane charge that also become unsuppressed in the $U\to \ii \infty$ limit. On the Type IIA side, these correspond to worldsheet instantons on curves $C$ that are linear combinations of $\mathbb{P}^1_f$ and $\mathbb{P}^1_b$. These can be computed in Type IIA using mirror symmetry but we do not discuss them here in detail. Instead, we stress that in the limit $U\to \ii \infty$, at constant $S,T$, on the one hand, D$(-1)$-instantons become unsuppressed that signal a breakdown of the perturbative $g_s$ expansion of Type IIB string theory which, however, is correctly taken into account by the geometry of the F-theory compactification. On the other hand, the D3-brane instantons are non-perturbative in both $g_s$ and $\alpha'$, indicating that also the geometric description of the theory breaks down.

\section{Details on \texorpdfstring{$\mathbb{P}^4_{2,2,2,3,3}[12]/(\mathbb{Z}_6\times\mathbb{Z}_2^2)$}{P22233[12]/(Z6xZ2xZ2)}}\label{app:CYex}
In this appendix we collect geometric details of the mirror of the resolved Calabi--Yau threefold $\mathbb{P}^4_{2,2,2,3,3}[12]$. In Section~\ref{ssec:oriendegenerations}, we consider complex structure degenerations of this mirror manifold. In order for the physical interpretation of~\cite{Hassfeld:2025uoy,Monnee:2025ynn} to hold, these degenerations must be semi-stable in the sense of Deligne--Mumford. By analysing the relevant singularities, we show that for this example this is indeed the case.

As a generic degree $12$ hypersurface $V'$ in $\mathbb{P}^4_{2,2,2,3,3}$ is quasi-smooth, its only singularities are those inherited from the ambient space~\cite{Fletcher:2000}.
The singular locus of the ambient space $\mathbb{P}^4_{2,2,2,3,3}$ consists of the curve $C'=\{x_1=x_2=x_3=0\}$ and the surface $S=\{x_4=x_5=0\}$. While the surface $S$ intersects the generic degree $12$ hypersurface $V'$ in a curve $C$, the curve $C'$ intersects this hypersurface in four points. Using the methods of~\cite{Schimmrigk:1989tz,Candelas:1989hd} we find that the singularities along $C$ can be resolved by a single blow-up of $C$, resulting in an exceptional divisor which is a $\mathbb{P}^1$-bundle over $C$ inside the hypersurface. Similarly, each of the four singular points in which $C'$ intersects the hypersurface can be resolved by a single blow-up, giving rise to four exceptional divisors $\mathbb{P}^2$. Using the simple criterion of~\cite{MR927963}, all singularities are seen to be Gorenstein, hence allowing for a crepant resolution~\cite{ROAN1996489}. Combined with the restriction of the hyperplane of $\mathbb{P}^4_{2,2,2,3,3}$, we find that the resolved Calabi--Yau $\Hat{V}$, corresponding to the proper transform of the zero locus
\begin{equation}\label{eq:P22233-app}
    \{P=x_1^6+x_2^6+x_3^6+x_4^4+x_5^4+\psi x_1x_2x_3x_3x_4x_5+\phi_1(x_1x_2x_3)^2+\phi_2(x_4x_5)^2+...=0\}\subset\mathbb{P}^4_{2,2,2,3,3}
\end{equation}
under the five blow-ups, has $h^{1,1}(\hat{V})=6$. As mentioned above, we are interested in complex structure degenerations of the mirror $V$ of the smooth Calabi--Yau $\hat{V}$. 

\subsection{Semi-stability of degenerations}
To show semi-stability of these, we have to determine the proper transform of~\eqref{eq:P22233-app} under the blow-ups. 

\paragraph{Proper transform and degenerations.} We will explicitly resolve the curve $C$ in $V'=\mathbb{P}^4_{2,2,2,3,3}[12]$ which consists of $\frac{1}{2}(1,1)$ (Gorenstein) singularities in the two normal directions on $V'$. Let $\pi:{\rm Bl}_C(V')\to V'$ denote the blow-up of $C$. Then the exceptional divisor $E=\pi^{-1}(C)$ is a $\mathbb{P}^1$-bundle over $C$ with fiber coordinates $[y_4:y_5]$ and the blow-up ${\rm Bl}_C(V')$ is defined by the proper transform of~\eqref{eq:P22233-app} subject to the relation $x_4y_5=x_5y_4$. Introducing the blow-up coordinate $t$ such that locally $E=\{t=0\}$ and $ty_{4,5}=x_{4,5}$, we find that the proper and total transform of~\eqref{eq:P22233-app} is given by
\begin{equation}
    \{x_1^6+x_2^6+x_3^6+t^4(y_4^4+y_4^4)+\psi t^2x_1x_2x_3y_4y_5+\phi_1(x_1x_2x_3)^2+\phi_2 t^4y_4^2y_5^2+...=0\}\,.
\end{equation}
Consider now the degeneration $\phi_2\to\infty$ within this blow-up\footnote{Notice that here we resolved $V'$ for fixed complex structure moduli. For any given degeneration, for example $\phi_2\to\infty$, one can also perform blow-ups of the whole {\it family} of Calabi--Yau threefolds as was done in Section~\ref{sec:OTypeAEx}. This amounts to also imposing the relation $\phi_2=t\chi_2$, which introduces another copy of the exceptional divisor in the total transform of the central fiber. The proper transform, which we are interested in, is unchanged.}, which has central fiber given by
\begin{equation}
   \{t^4y_4^2y_5^2=0\}=\{y_4^2y_5^2=0\}+4E\,.
\end{equation}
This coincides with the total transform of the central fiber of the degeneration $\phi_2\to\infty$ in the unresolved space. As we are interested in the proper transform, however, we have to subtract from this total transform the exceptional contribution.\footnote{For a submanifold $Y\subset X$ containing the blow-up center $Z=C\cup\{p_{2,3}\}$, we use here the relation $\pi^{-1}(Y)=\Tilde{Y}+aE$ between the total and proper transform of $Y$ under the blow-up $\pi:{\rm Bl}_Z(X)\to X$, see e.g.~\cite{Iitaka}.} Thus,
\begin{equation}\label{eq:central-fiber-proper}
    {\rm Bl}_C(V')_0=\{y_4^2y_5^2=0\}\,,
\end{equation}
which is of the same functional form as the central fiber of the degeneration in the unresolved space. Blowing-up the four points in $V'\cap C'$ gives rise to similar relations. The upshot of this is that the normal crossing property of the central fiber can be checked at the level of the unresolved hypersurface~\eqref{eq:P22233-app} as the blow-up effectively acts as a relabelling of the coordinates. Notice that~\eqref{eq:central-fiber-proper} is not yet of simple normal crossing type as the two components are non-reduced. When shifting attention to the mirror $V$ of the fully resolved $\Hat{V}$, this non-reducedness is taken care of by the Greene--Plesser group to which we turn next.

\paragraph{Greene--Plesser construction.} The mirror $V$ of $\hat{V}$ can be found using the construction of~\cite{Greene:1990ud}. For this we focus on the Fermat part of $P$ and see that its naive symmetry group is given by $G_0=\mathbb{Z}_6^3\times\mathbb{Z}_4^2$. Using the notation of~\cite{Lust:2022mhk}, the subgroup of $G_0$ compatible with the Calabi--Yau condition of $\{P=0\}$ is given by
\begin{equation}
    G_{\rm CY}=\ker\big(\phi:(a_1,a_2,a_3,a_4,a_5)\mapsto 2(a_1+a_2+a_3)+3(a_4+a_5)\in\mathbb{Z}_{12}\big)\,,
\end{equation}
where $a_{1,2,3}\in\mathbb{Z}_6$ and $a_{4,5}\in\mathbb{Z}_4$. The order of this kernel is therefore $|G_0|/|{\rm im}(\phi)|=3456/12=288$. Generators of $G_{\rm CY}$ are given by
\begin{equation}
    g^{(1)}=(1,-1,0,0,0),\quad g^{(2)}=(1,0,-1,0,0),\quad g^{(3)}=(0,0,0,1,-1),\quad g^{(4)}=(3,0,0,-2,0)\,,
\end{equation}
where $g^{(4)}$ squares to the identity. Thus,
\begin{equation}
    G_{\rm CY}=\mathbb{Z}_6^2\times\mathbb{Z}_4\times\mathbb{Z}_2\,,
\end{equation}
with the $i$th-factor generated by $g^{(i)}$. To arrive at the Greene--Plesser group, we have to remove the ambient rescaling invariance $\langle R\rangle=\mathbb{Z}_{12}$ from $G_{\rm CY}$. Taking this into account, we find the additional relations
\begin{equation}
    (g^{(1)}g^{(2)})^2=(0,0,0,0,0)\,,\quad g^{(1)}g^{(2)}g^{(3)}=g^{(4)}\,,\quad (g^{(3)})^2=(0,0,0,0,0)\,,
\end{equation}
with the first one breaking one $\mathbb{Z}_6$-factor in $G_{\rm CY}$ to a $\mathbb{Z}_2$, the second relation removing the $\mathbb{Z}_2$-factor from $G_{\rm CY}$ and the third breaking $\mathbb{Z}_4$ to another $\mathbb{Z}_2$. Hence,
\begin{equation}
    G_{\rm GP}=\mathbb{Z}_6\times\mathbb{Z}_2^2\,,
\end{equation}
with $|G_{\rm GP}|=24=|G_{\rm CY}|/|\langle R\rangle|$. There are two sets of generators that will be convenient for the F-theory lifts we study in Section~\ref{ssec:Ftheorylift}. For the $x_1\leftrightarrow x_2$ orientifold, we consider
\begin{equation}\label{eq:P22233-12-GGP}
    g^{(1)}=(1,-1,0,0,0)\,,\quad\Tilde{g}^{(2)}=(1,1,-2,0,0)\,,\quad g^{(3)}=(0,0,0,1,-1)\,,
\end{equation}
while for the $x_4\leftrightarrow x_5$ orientifold we replace $g^{(3)}$ by
\begin{equation}\label{eq:P22233-45-GGP}
    \Tilde{g}^{(3)}=(3,0,0,1,1).
\end{equation}
The three non-Fermat terms written explicitly in~\eqref{eq:P22233-app} are the only monomials invariant under $G_{\rm GP}$, which seems to be in conflict with $h^{2,1}(V)=h^{1,1}(\hat{V})=6$. Notice, however, that the Calabi--Yau $\mathbb{P}^4_{2,2,2,3,3}[12]$ is non-favourable~\cite{Anderson:2008uw}, meaning there are (in this case three) non-toric divisors on the hypersurface which do not arise from ambient space divisors. On the mirror side, these K\"ahler moduli of $\hat{V}$ map to non-polynomial complex structure deformations of $V$. In the analysis of Sections~\ref{ssec:oriendegenerations} and~\ref{ssec:Ftheorylift} we are therefore considering only a subspace of the full complex structure moduli space of $V$. When taking an extreme limit for one of the moduli $\psi,\phi_{1,2}$, we assume the other five deformation parameters to be fixed at generic values.\\

Notice that taking the quotient by $G_{\rm GP}$ introduces new orbifold singularities so that we do not end up with a smooth mirror Calabi--Yau. By definition of $G_{\rm GP}$, all group elements respect the Calabi--Yau condition, which means that all singularities introduced by $G_{\rm GP}$ are locally of the form $\mathbb{C}^3/G$ with $G\subset{\rm SL}(3,\mathbb{C})$ an abelian subgroup. In other words~\cite{MR927963}, all new singularities are Gorenstein and hence allow for a crepant resolution~\cite{ROAN1996489}. As these resolutions only affect the K\"ahler sector of $V$, for the purpose of considering limits in the complex structure moduli space of $V$, their specifics are not of importance.

\paragraph{Semi-stability of threefold degenerations.} Around~\eqref{eq:central-fiber-proper} it was shown that blowing-up singularities does not alter the functional form of the central fiber of a complex structure degeneration of a Calabi--Yau threefold. However, to conclude that this degeneration is semi-stable in the resolved space, it is still to show that the components into which the resolved threefold splits, are reduced. Using the Greene--Plesser group computed in the previous paragraph, we see that on the mirror $V$ of $\Hat{V}=\mathbb{P}^4_{2,2,2,3,3}[12]$ the components of the central fibers in the limits $\psi,\phi_{1,2}\to\infty$ are indeed reduced, showing that these degenerations are semi-stable. 

\paragraph{Semi-stability of fourfold degenerations.} In Section~\ref{ssec:Ftheorylift} we construct the F-theory lifts of two orientifold projections of $V$. While we refer to the main text for the explicit construction, we discuss here how the singularities of the ambient space of $\cB_3$ have changed compared to those of $\mathbb{P}^4_{2,2,2,3,3}$. From the fourfold perspective, all degenerations considered in Section~\ref{ssec:Ftheorylift} are degenerations of the base $\cB_3$. Therefore it is enough to analyse the singularity structure on the base to determine whether the degeneration is semi-stable or not. Formally, we are therefore in the same situation as in the Calabi--Yau threefold case discussed before. Adapting the previous analysis to the F-theory base $\cB_3$ shows that its degenerations studied in Section~\ref{ssec:Ftheorylift} are semi-stable. Therefore, the same conclusion holds also from the fourfold perspective. 

\subsection{F-theory uplift of an O-Type A Type II limit}\label{sapp:TypeOAEx}
In this appendix we collect details on the generic Weierstrass model over the base $\cB_{3,0}$ considered in Section~\ref{sec:OTypeAEx}. Recall from Section~\ref{ssec:oriendegenerations} that $\cB_3$ is the base of the F-theory uplift of the orientifold of $\mathbb{P}^4_{2,2,2,3,3}[12]/G^\sigma_{\rm GP}$ defined by the exchange $\sigma:x_4\leftrightarrow x_5$. Here, $G^\sigma_{\rm GP}$ denotes the subgroup of the Greene--Plesser group $G^\sigma_{\rm GP}$ invariant under the action of $\sigma$. For this choice of $\sigma$ one finds $G^\sigma_{\rm GP}=G_{\rm GP}=\mathbb{Z}_6\times\mathbb{Z}_2^2$. 

The base $\cB_3$ of the F-theory fourfold $W$ is given as a degree $12$ hypersurface inside $\mathbb{P}^4_{2,2,2,3,6}/G^\sigma_{\rm GP}$. In the complex structure limit $\tilde{\phi}_2\to\infty$, the base degenerates as
\begin{equation}
    \cB_3\longrightarrow\cB_{3,0}=\{(p^2-k)^2=0\}\,,
\end{equation}
where $[x_1:x_2:x_3:p:k]$ are the coordinates of the ambient space of $\cB_3$. The generators of $G^\sigma_{\rm GP}$, as given in~\eqref{eq:P22233-12-GGP} and~\eqref{eq:P22233-45-GGP}, act on this ambient space as (dropping the tilde on $\Tilde{g}^{(2)}$ and $\Tilde{g}^{(3)}$)
\begin{align}
g^{(1)}:\ (x_1,x_2,x_3,p,k) &\longmapsto (a\,x_1,\ a^{-1}x_2,\ x_3,\ p,\ k)\,,\\
g^{(2)}:\ (x_1,x_2,x_3,p,k) &\longmapsto (a\,x_1,\ a\,x_2,\ a^{-2}x_3,\ p,\ k)\,,\\
g^{(3)}:\ (x_1,x_2,x_3,p,k) &\longmapsto (a^{3}x_1,\ x_2,\ x_3,\ c\,p,\ d\,k)\,,
\end{align}
where $a=e^{2\pi i/6}$, $c=e^{2\pi i/4}=i$, and $d=e^{2\pi i/2}=-1$.

For the analysis in Section~\ref{sec:OTypeAEx} we need the most general form of the functions $f$ and $g$ appearing in the Weierstrass polynomial for a Weierstrass model over the degenerate base $\cB_{3,0}$. Recall that $\bar{K}_{\cB_3}$ has degree $3$ so that $\deg(f)=12$ and $\deg(g)=18$ in the base coordinates. To find the most general $G_{\rm GP}$--invariant functions $f$ and $g$, consider a monomial
\begin{equation}
m=x_1^{e_1}x_2^{e_2}x_3^{e_3}p^{e_p}k^{e_k}
\end{equation}
and observe that the invariance conditions under $g^{(1)},g^{(2)}$ are
\begin{equation}
    e_1-e_2\equiv 0\pmod 6\,,\qquad e_1+e_2-2e_3\equiv 0\pmod 6\,,
\end{equation}
while invariance under $g^{(3)}$ requires
\begin{equation}\label{eq:g3condition}
(-1)^{e_1} i^{e_p}(-1)^{e_k}=1\,.
\end{equation}
In particular, $e_p$ must be even, say $e_p=2r$, so that \eqref{eq:g3condition} implies
\begin{equation}
    r+e_1+e_k\equiv 0\pmod 2\,.
\end{equation}
Imposing the base relation $k=p^2$ eliminates $k$ in favour of $p$, so that $f$ and $g$ generically involve all degree $12$ and $18$ $G_{\rm GP}$--invariant polynomials in the coordinates $p,x_1,x_2,x_3$, which are
\begin{equation}\label{eq:P22233-34-generic-f}
    \begin{aligned}
    f=\;& a_0\,x_1^6+a_1\,x_2^6+a_2\,x_3^6+a_3\,(x_1x_2x_3)^2+a_4\,x_1x_2x_3p^2+a_5\,p^4
    \end{aligned}
\end{equation}
and
\begin{equation}\label{eq:P22233-34-generic-g}
    \begin{aligned} g=\;&b_0\,x_3^9+b_1\,x_1^6x_3^3+b_2\,x_2^6x_3^3+b_3\,(x_1x_2)^4x_3+b_4\,x_1^2x_2^2x_3^5\\
    &\qquad+b_5\,x_3^3p^4+b_6\,(x_1x_2)^3p^2+b_7\,x_1x_2x_3^4p^2\,.
    \end{aligned}   
\end{equation}
For generic non-vanishing coefficients $a_i$, $b_j$, the resulting Weierstrass model is smooth. To see this, recall that according to the classification of Kodaira, the elliptic fiber becomes singular over those loci where the discriminant $\Delta$ vanishes.  As long as the vanishing orders of $(f,g, \Delta)$  do not reach or exceed the values $(4,6,12)$ in codimension one on the base, the model is called minimal, and the type of singularity in the fiber can then be read off from the specific values of the vanishing orders. For example, vanishing orders ${\rm ord}(f,g, \Delta) = (0,0,n)$ indicate a Kodaira fiber I$_n$, at which $g_s \to 0$. By contrast, at singularities where also $f$ and $g$ vanish, $g_s$ typically takes non-perturbative values. For the precise statement and more details, we refer to Table  4.1 in \cite{Weigand:2018rez} and the original references therein.

\bibliography{papers_Max}
\bibliographystyle{JHEP}

\end{document}